\documentstyle[pre,aps,multicol,epsf]{revtex}
\begin{document}

\title{Hydration interactions: aqueous solvent effects in electric double layers\thanks{Accepted to {\it Phys. Rev. E.}}}
\author{Yoram Burak\thanks{email: yorambu@post.tau.ac.il} and David Andelman\thanks{email: andelman@post.tau.ac.il}}
\address{School of Physics and Astronomy, Raymond and Beverly Sackler
  Faculty of Exact Sciences, \\ Tel Aviv University,
Tel Aviv 69 978, Israel}
\date{5 July 2000}
\maketitle


\begin{abstract}

A model for ionic solutions with an 
attractive short-range pair
interaction between the ions is presented. The short-range
interaction is accounted for by adding a quadratic non-local term
to the Poisson-Boltzmann  free energy. The model is used to study
solvent effects in a planar electric double layer.
The counter-ion density is found to increase near the charged surface,
as compared with the Poisson-Boltzmann theory, 
and to decrease at larger distances.
The ion density profile is studied
analytically in the case where the ion distribution near the plate
is dominated only by counter-ions.  Further away from the plate the
density distribution can be described using a Poisson-Boltzmann theory 
with an effective surface charge that is smaller than the actual one.
\end{abstract}


\begin{multicols}{2}

\section{Introduction}
\label{sec:intro}

Electrolytes, in contact with charged surfaces or macro-ions, play
an important role in determining the properties of many biological
and chemical systems. One of the most widely used tools for
studying ions in aqueous solutions is the Poisson-Boltzmann (PB)
theory \cite{Gouy10,Chapman13,DH23,DH24,Israelachvili,Andelman95}. 
The mathematical and
conceptual simplicity of this theory makes it very appealing both
for numerical computation\cite{HonigNicholls95}, and for gaining insight
into the underlying physical principles. Although the theory
contains important simplifications, it has proven to be a useful
and accurate tool in the study of systems such as
colloidal dispersions\cite{VO48,ACGMPH84}, 
biological membranes\cite{Andelman95}, 
polyelectrolytes\cite{BarratJoanny96} 
and complex systems, {\it e.g.}, proteins or DNA interacting 
with charged membranes\cite{HMGB98,BAO98,FleerStuart}.

The Poisson-Boltzmann theory is obtained by making two simplifying
approximations. The first approximation is the treatment of the
electrostatic interactions on a mean-field level. The ions are
treated as independent charged particles interacting with an
external electrostatic potential, derived self-consistently from
the mean charge density distribution. Thus, correlations between
the ion positions are not taken into account. The second
approximation is the treatment of the ions as point-like objects,
interacting only through the electrostatic interaction in a
dielectric medium. In reality, ions in aqueous solutions have more
intricate interactions \cite{Israelachvili}. These include a
non-Coulombic interaction between ion pairs, which is mainly a
short-range steric repulsion, interactions with the polar solvent
molecules and short-range interactions with the confining charged
surfaces.

Various models have been proposed for the inclusion of effects not
accounted for by the PB theory. These include liquid state theory
approaches
\cite{KjellanderMarcelja85,Kjellander88,LOB81,OuthBhu83,BlumHenderson92},
field theory expansions \cite{NetzOrland99EPL,NetzOrland00},
computer simulations \cite{GJWL84,KAJM92,GKA97} and other
modifications to the PB theory
\cite{StevensRobbins90,BAO97,BAO00,LukSaf99}. Most of these models
remain within the framework of the so-called ``primitive model'',
in which the interaction between the ions is modeled as a purely
repulsive hard-core interaction. On the other hand, relatively few
works have addressed explicitly the discrete nature of the solvent
molecules
\cite{BlumHenderson92,TorPat91,IsraelWenner96,OttoPat99}. Clearly,
the replacement of the solvent by a continuous medium cannot be
precise when the inter-ion distance is comparable to
the solvent molecular size. Therefore, when the ions reach high
densities the discreteness of the solvent is expected to have an
important effect on the ionic distribution. This is of particular
importance for water. Due to its high polarity, the strong
screening of the electrostatic interaction (represented by the
dielectric constant) is modified at small
 ion separations.

Using the surface force apparatus\cite{Israelachvili}, 
it is possible to 
measure precisely the force between charged mica plates. 
These measurements supply 
evidence for the importance of the solvent structure
in aqueous solutions
\cite{Pashley81a,PashIsrael84}.
At inter-plate separations below approximately 
$20\,{\mbox \AA}$ significant deviations are found 
from the prediction of the 
Derjaguin-Landau-Verwey-Overbeek (DLVO) theory 
\cite{VO48,DL41}. The measured
force is oscillatory or consists of a series of
steps, with a period corresponding to the water molecular size.
Oscillatory forces are known to arise as a result of the solvent
structuring in layers between surfaces \cite{Israelachvili}.
However, a repulsive contribution is found in addition to the
oscillatory force at plate separations below several nanometers
\cite{Pashley81a,PashIsrael84}.
This repulsive force is often referred to as the ``hydration
force'' \cite{Israelachvili,Pashley81a},
and its origin is not yet completely understood \cite{IsraelWenner96}.

\subsection{Aqueous pair potential model}

Recently \cite{Marcelja97Nature,Marcelja99}, an aqueous pair
potential model has been proposed for electrolytes, in which the
effect of the solvent on the ions is described as a short-range
two-body interaction between the ions. The solvent is replaced by
a continuum dielectric medium as in PB theory, but the ions also
interact through a two-body short-range hydration interaction
\cite{Marcelja97Nature}. This is shown schematically in Fig.~1.

This aqueous pair potential model \cite{Marcelja97Nature,Marcelja99}, 
involves several simplifying
assumptions. One is that the effect of the solvent can be
represented as a linear superposition of two-body potentials
between all ion pairs. Another simplification is that the
effective potential between the ions is taken as the effective
potential in the bulk, regardless of the ion concentration, and of
the geometry imposed by the charged surfaces. 
Finally, a short-range surface-ion
effective potential should be included in addition to the ion-ion
effective potential.  Despite of the simplifications made in the
aqueous pair potential model, it offers a first step
towards a
qualitative understanding of solvent effects on the ion
distribution, in particular near highly charged surfaces.

\subsection{Effective ion pair interaction}

For the short-range ion-ion interaction, the so-called potential
of mean force between ions in solution can be used. Potentials of
mean force are defined as $-k_{\rm B}T\log g_{ij}({\bf r})$ where
$g_{ij}({\bf r})$ are the ion-ion radial distribution functions
for ion pairs of species $i$ and $j$. The radial distribution
functions have been calculated numerically for a single ion pair
immersed in an aqueous solution using molecular dynamics
techniques \cite{GRP91a,GRP91b,DangSmith94}.

An alternative approach has been proposed in Refs.
\cite{Lyubartsev97,Lyubartsev95}. In this approach, a Hamiltonian
consisting of a pairwise effective potential between the ions is
obtained using the so-called ``reverse Monte-Carlo'' approach. The
ion-ion radial distribution functions are first calculated using a
molecular dynamics simulation for a system including solvent
molecules and a finite concentration of ions.
The ion-ion effective potential in the system without
the solvent is then adjusted iteratively until the same
distribution functions are obtained using Monte-Carlo simulations.

The different available calculations of potentials of mean force
differ in their quantitative predictions. This may be a result of
high sensitivity of the models to detailed features used for the
water molecules and for the inter-molecular interactions
\cite{GRP91a}. However, all the potentials of mean force as 
well as
the effective potentials ~\cite{Lyubartsev97}
are qualitatively similar \cite{Marcelja98}. Thus,
for the purpose of the present work, aiming at a
qualitative understanding of solvent effects, any one
of these potentials may be used.

\end{multicols}
\begin{figure}[b]
\epsfxsize=0.35\linewidth \centerline{\hbox{
\epsffile{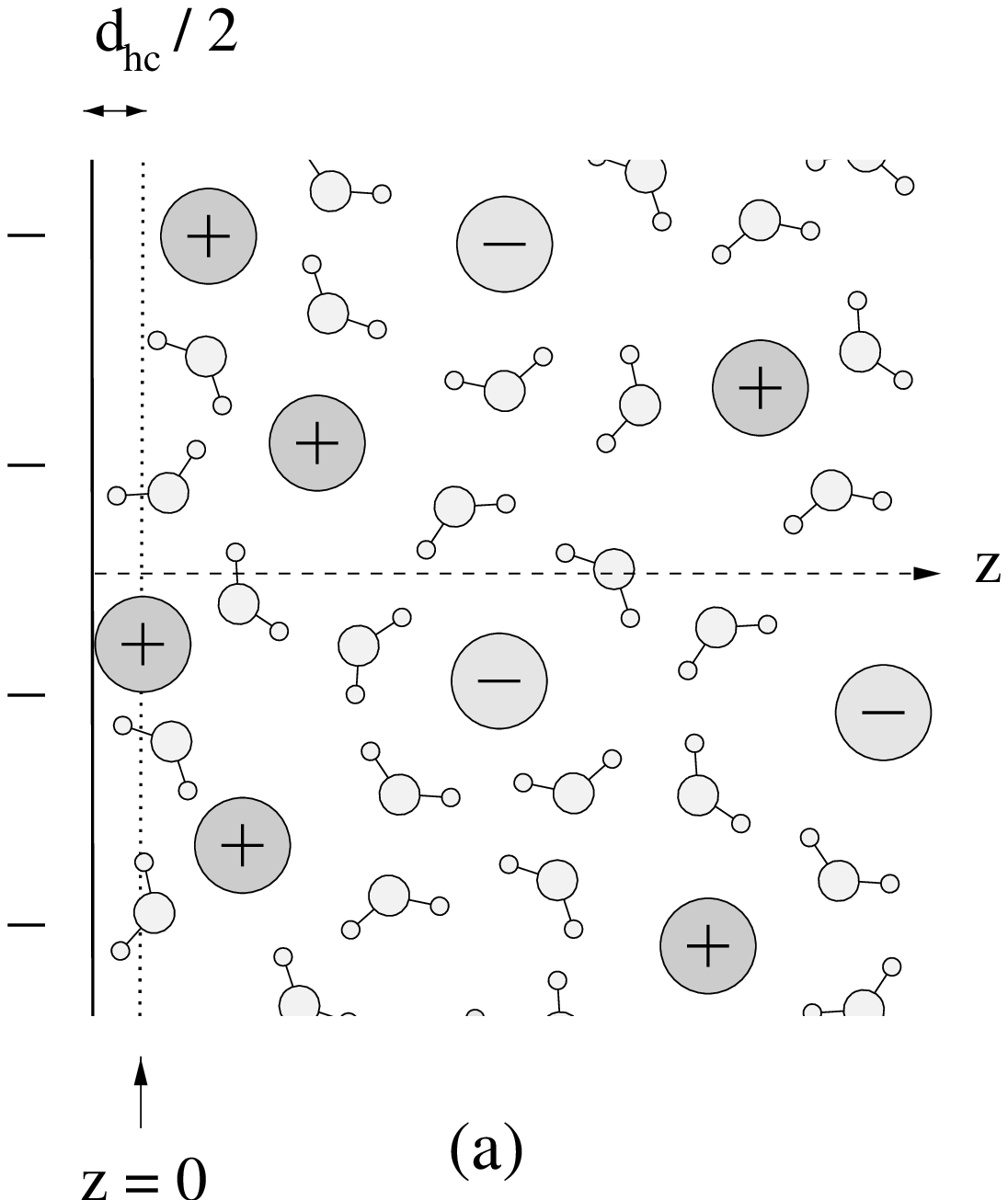} } \hspace{0.05\linewidth}
\epsfxsize=0.35\linewidth
        \hbox{ \epsffile{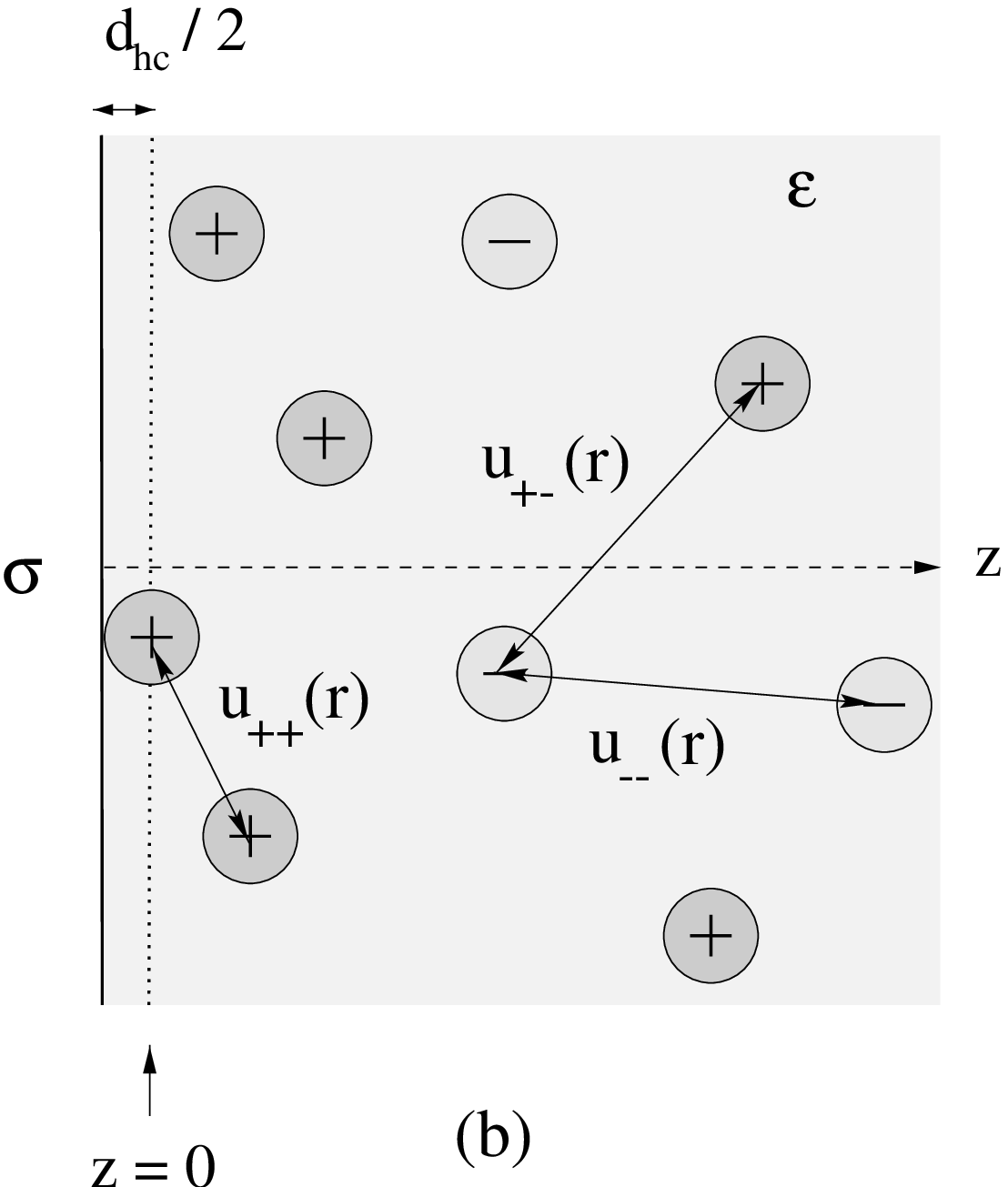} } }
\caption{Schematic description of
the aqueous pair potential model. An aqueous ionic solution
in contact with a charged plate in (a) is replaced 
in (b) by ions
in a continuum dielectric medium having a dielectric
constant $\varepsilon$, with electrostatic and
short-range interactions 
$u_{ij}(r) = u_{ij}(|{\bf r}|)$.
The $z$ coordinate designates the distance from the
charged plate, with $z = 0$ corresponding to the
distance of closest approach of the ions to the plate.
The distance of closest approach is equal to $d_{\rm hc}/2$,
where $d_{\rm hc}$ is the hard-core diameter of the ions.
}
\end{figure}
\begin{multicols}{2}

\begin{figure}[tbh]
\epsfxsize=0.9\linewidth \centerline{\hbox{
\epsffile{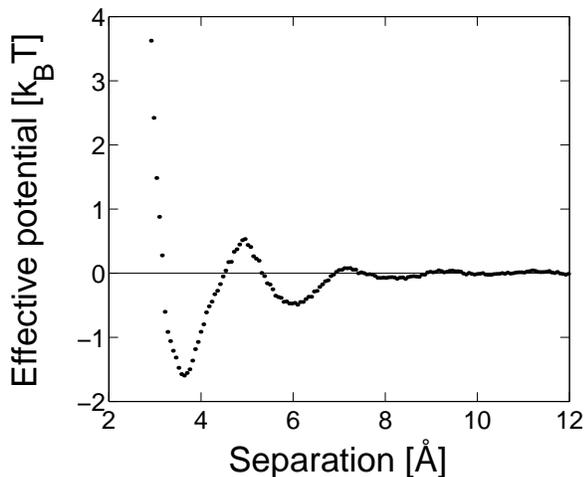} } }
\caption{Short-range effective potential between
\protect{${\rm Na}^{+}$} ion pairs, adapted from 
Ref.~\protect\cite{Lyubartsev97}
using simulations in a bulk NaCl solution of concentration 
$0.55$\,M, at room temperature 
\protect\cite{Marcelja00pr}. 
The potential is shown in units of 
$k_{\rm B}T$, as a
function of the distance between the ion centers. For ion
separations smaller than $2.9\,\mbox{\AA}$ 
a hard core interaction was
assumed. The Coulomb interaction is subtracted to show only the
short-range hydration effect due to the water molecules.
}
\end{figure}

At large ionic separations the ion-ion effective potential is well
approximated by a screened electrostatic interaction, with the
water dielectric constant in the continuum limit. At short ionic
separation, the difference between the total effective potential
and the screened electrostatic interaction is a short-range
potential reflecting the structure of water molecules in the ion
vicinity. Fig.~2 shows the short-range contribution (excluding the
screened electrostatic part) to the effective
potential calculated between \mbox{${\rm Na}^{+}$ - ${\rm
Na}^{+}$} pairs in the reverse Monte-Carlo approach
\cite{Lyubartsev97}. Below
about $3$\,\AA, the electrostatic repulsion between the ions
becomes unscreened. Therefore it is much larger than the
screened repulsion in the dielectric medium, 
and the effective potential is
strongly repulsive. The unscreened electrostatic potential leads
to an effectively enlarged hard-core
separation between the ions, relative to a hard-core diameter
of about $2.3\,\mbox{\AA}$ used in the short-range part of
the bare ion-ion potential. At larger separations, the effective
potential is oscillatory, and mainly attractive. It has a distinct
minimum at an ion-ion separation of about $3.6$\,\AA,
followed by a maximum and a second minimum at
approximately $6$\,\AA.

\subsection{The present work}

The replacement of the discrete solvent by a continuum medium,
with electrostatic and short range interactions between the 
ions, is a considerable simplification.
Still, the statistical mechanical treatment of an
electrolyte solution in this 
model is difficult, and requires 
the use of further approximations, or simulations.

The Anisotropic Hyper-Netted Chain approximation (AHNC) 
\cite{Kjellander88} was previously used
to calculate the effects of hydration interactions in the aqueous
pair potential model~\cite{Marcelja98,Marcelja97}.
When the ion concentration is large enough, {\it e.g.}, near a
highly charged surface, the hydration interaction is found to have
a significant effect on the distribution of ions in the solution.
It was also proposed that the so-called repulsive
 ``hydration forces'' between surfaces arise
from the ionic structure near highly charged surfaces. According
to this description, at large distances from the plate, the ion
distribution follows a PB profile with a reduced effective surface
charge. When two plates approach each other, the ions near the
surfaces come into contact giving rise to an apparent new
repulsive force.

In the present work  a simple description for ions interacting
through electrostatic and short-range attractive interactions as
mediated by the solvent molecules is introduced. 
We apply this description to
the aqueous pair potential model. Our aim is limited to describe
the important effects of the short-range interaction, and not to
provide an accurate tool for their calculation. Therefore our
model follows the PB theory as closely as possible, and describes
the short-range interaction using a simplified term added
to the free energy. The advantage of this approach over more
elaborate treatments such as the AHNC
\cite{Marcelja98,Marcelja97}, is that it
provides relatively simple equations that can be treated
numerically and analytically with relative ease as well as allowing
extensions to non-planar geometries.
In a planar geometry, we show that the effect of the
ion-ion hydration interaction can be understood as a perturbation
over the PB results. 
An increase in the concentration of counter-ions near the 
charged surface is found, and it results in
an apparent surface charge which is reduced relatively to the PB
theory.

In addition to the calculation of charge distributions, the
effect of the hydration interaction on the force between charged
particles or surfaces can be studied and will be presented 
elsewhere~\cite{BurakAndelman00b}. Since we will not discuss
inter-surface forces in this paper, it is worthwhile to mention
that the results obtained using our model are in good agreement
with AHNC calculations \cite{Marcelja00pr} and may
provide an explanation to the hydration forces as observed in
surface force measurements \cite{Pashley81a,PashIsrael84}.
For high surface charge and plate
separations up to approximately $20\,\mbox{\AA}$, important
modifications of the PB predictions are found. 

The outline of the paper is as follows. Section \ref{sec:model}
presents the model. In section \ref{sec:oneplate} we apply the
model to a single charged plate, present numerical results for the
ion density profile and discuss the modifications to the PB
theory due to the addition of short-range interactions. In section
\ref{sec:analytical} we present analytical results in the
low salt limit. We calculate the effective PB surface charge and
the effect of the hydration interaction on the density profile of
counter-ions in a system with no added salt.

\section{The Model}
\label{sec:model}

\subsection{Free energy}

We start from an approximated free energy, written as a functional
of the various ion densities. We choose the electrostatic boundary
conditions to be of fixed surface charge densities and write the
free energy as a sum of the usual PB term and a
correction term, due to hydration, as will be explained below:

\begin{equation}
\label{eq:Omega} \Omega  =  \Omega_{\rm PB} + \Delta \Omega \\
\end{equation}

\noindent We discuss first how the PB free energy
is obtained, and then generalize this result to include
the short-range hydration interaction.

\subsubsection{Poisson-Boltzmann free energy}

 The Hamiltonian of the system is:

\begin{equation}
H = \frac{1}{2}\sum_{i}\int_{V}{\rm d}^{3}{\bf r}\,
        e_{i}\rho_{i}({\bf r}) \phi({\bf r}) +
    \frac{1}{2}\oint_{\partial V}{\rm d}^{2}{\bf r}_{\rm s}\,
        \sigma({\bf r}_{\rm s}) \phi({\bf r}_{\rm s})
\label{eq:H}
\end{equation}

\noindent where $V$ is the the volume occupied by the electrolyte
solution, $\sigma({\bf r}_{\rm s})$ is the surface charge density
of immobile charges on the 
boundaries $\partial V$, and $e_{i}$ is the charge of the
$i$th ion species. The ion densities $\rho_{i}({\bf r})$ are:

\begin{equation}
\rho_{i}({\bf r}) \equiv \sum_{j=1}^{N_{i}}
              \delta({\bf r}-{\bf r}_{i}^{j})
\end{equation}

\noindent where ${\bf r}_{i}^{j}$ is the position of the $j$th ion
of the $i$th species, and 
the electrostatic potential $\phi({\bf r})$ is
a function the different ion positions:

\begin{equation}
\phi({\bf r})= \sum_{j}\int_{V}{\rm d}^{3}{\bf r'}\,
   \frac{e_{j}\rho_{j}({\bf r'})}
            {\varepsilon \left|{\bf r}-{\bf r'}\right|}
    + \oint_{\partial V}{\rm d}^{2}{\bf r}_{\rm s}\,
   \frac{ \sigma({\bf r}_{\rm s})}
            {\varepsilon \left|{\bf r}-{\bf r}_{\rm s}\right|}
\label{eq:phi}
\end{equation}

\noindent where $\varepsilon =78$ is the dielectric constant
of water. The PB theory is obtained by using a mean-field
approximation for the electrostatic interaction. The Hamiltonian
(\ref{eq:H}) is first replaced by a mean-field Hamiltonian, where
the electrostatic potential $\phi({\bf r})$ is replaced by an
external field $\Psi({\bf r})$. In the thermodynamic limit the
free energy can then be written as a functional of the mean
densities of the ion species, $c_{i}({\bf r}) = \langle
\rho_{i}({\bf r})\rangle_{\rm MF}$, as follows:
\begin{eqnarray}
\Omega_{\rm MF} & = & k_{B}T
   \int_{V} \sum_{i} c_{i}
     \left[ \log \frac{c_{i}}{\zeta_{i}} - 1 \right]
   {\rm d}^{3}{\bf r} \nonumber \\
& & + \frac{1}{2}
   \int_{V} \sum_{i} e_{i}\Psi({\bf r})c_{i}({\bf r})
   {\rm d}^{3}{\bf r}
\nonumber \\
& & + \frac{1}{2} \oint_{\partial V}\sigma({\bf r}_{s})
     \Psi({\bf r}_s)
\label{eq:Omegaex}
\end{eqnarray}

\noindent where $k_{\rm B}T$ is the thermal energy, and
$\zeta_{i}$ is the fugacity of the $i$th ion species.
The mean-field approximation is
obtained by requiring that the external potential $\Psi({\bf r})$
is the thermodynamical average of (\ref{eq:phi}) in the system with
the mean-field Hamiltonian, $\Psi({\bf r}) = \langle \phi({\bf r})
\rangle_{\rm MF}$, {\it i.e.}:

\begin{equation}
\Psi({\bf r})=
   \sum_{i}\int_{V}{\rm d}^{3}{\bf r'}\,
   \frac{e_{i}c_{i}({\bf r'})}
            {\varepsilon \left|{\bf r}-{\bf r'}\right|}
    + \oint_{\partial V}{\rm d}^{2}{\bf r}_{\rm s}\,
   \frac{\sigma({\bf r}_{\rm s})}
            {\varepsilon \left|{\bf r}-{\bf r}_{\rm s}\right|}
\label{eq:phiex}
\end{equation}

\noindent This relation is equivalent to the Poisson equation:

\begin{equation}
{\bf \nabla}^{2}\Psi 
= -\frac{4\pi}{\varepsilon}\sum_{i}e_{i}c_{i}
\label{eq:Poisson}
\end{equation}

\noindent supplemented by the boundary condition:

\begin{equation}
\left. {\bf \nabla} \Psi \cdot \hat{\bf n}\right|_
{\textstyle{\bf r_s}} 
= - \frac{4\pi}{\varepsilon}\sigma({\bf r_s})
\ \ \ \ \ \mbox{on the charged surfaces}\label{eq:boundary}
\end{equation}

\noindent 
where the normal vector $\hat{\bf n}$ points away from the
charged surfaces into the volume occupied by the ionic solution.
Using this boundary condition and Eq.~(\ref{eq:Poisson}),
the second and third terms of Eq.~(\ref{eq:Omegaex}) can be
re-expressed as:

\begin{eqnarray}
\frac{1}{2}
   \int_{V} \sum_{i} e_{i}\Psi({\bf r})c_{i}({\bf r})
   {\rm d}^{3}{\bf r}
   + \frac{1}{2} \oint_{\partial V}\sigma({\bf r}_{s})
     \Psi({\bf r})
\nonumber \\ =
    \frac{\varepsilon}{8\pi} \int_{V} ({\bf \nabla} \Psi)^{2}
        {\rm d}^{3}{\bf r}
\end{eqnarray}

\noindent Substituting this relation in Eq.~(\ref{eq:Omegaex}) we
obtain the PB free energy:

\begin{eqnarray}
\Omega_{\rm PB} & = &  \frac{\varepsilon}{8\pi}
   \int ({\bf \nabla} \Psi)^{2}
   {\rm d}^{3}{\bf r} 
\nonumber \\
 & + & k_{B}T
   \int \sum_{i} c_{i} \left[ \log
          \frac{c_{i}}{\zeta_{i}} - 1 \right]
   {\rm d}^{3}{\bf r}
\nonumber \\
 & + & \int \lambda ({\bf r}) \left(
   {\bf \nabla}^{2}\Psi + \frac{4\pi}{\varepsilon}\sum_{i}e_{i}c_{i}
                              \right)
   \,{\rm d}^{3}{\bf r}
\label{eq:OmegaPB}
\end{eqnarray}

\noindent 
The first term in $\Omega_{\rm PB}$ is the electrostatic free
energy and the second term is the entropy of the ions. 
The fugacity $\zeta_{i}$, in the second term, is equal 
in PB theory to the bulk concentration $c_{{\rm b},i}$ 
of the $i$th ion species, $\zeta_{i}=c_{{\rm b},i}$,
as for an ideal gas.
For more generalized free energies, 
a different relation may exist between
the fugacity of each ion species and its respective 
bulk concentration. The
electrostatic potential $\Psi$ is a functional of the ion
densities $c_{i}$, and is determined by the Poisson
equation~(\ref{eq:Poisson}) and the boundary conditions
(\ref{eq:boundary}) imposed by the surface charges. Alternatively,
in Eq.~(\ref{eq:OmegaPB})
$\Psi$ is regarded as an independent field and a
third term containing a Lagrange
multiplier $\lambda({\bf r})$ is added to $\Omega_{\rm PB}$.
The PB equilibrium mean densities
$c_{i}({\bf r})$ result from minimizing
$\Omega_{\rm PB}$. With the introduction of
$\lambda({\bf r})$ the minimization is equivalent to
requiring an extremum of $\Omega_{\rm PB}$ with respect to
the three fields $c_{i}$, $\Psi$ and $\lambda$, subject to the
boundary condition (\ref{eq:boundary}). By requiring first 
an extremum of  $\Omega_{\rm PB}$ with
respect to $\Psi$ and $c_{i}$ the following 
relations are obtained:

\begin{equation}
\lambda = \frac{\varepsilon}{4\pi}\Psi \label{eq:min1}
\end{equation}

\noindent and:

\begin{equation}
c_{i} =\zeta_{i}\exp\left(-\beta e_{i} \Psi \right)
\label{eq:min2}
\end{equation}

\noindent  
Where $\beta = 1/(k_{\rm B}T)$ and $\zeta_i=c_{{\rm b},i}$. 
The extremum condition with respect to $\lambda$ gives the
Poisson equation:

\begin{equation}
{\bf \nabla}^{2}\Psi =-\frac{4\pi}{\varepsilon}\sum_{i}e_{i}c_{i}
\end{equation}

\noindent Combining these relations we obtain the PB equation:

\begin{equation}
{\bf \nabla}^{2}\Psi = -\frac{4\pi}{\varepsilon}
                  \sum_{i}\zeta_{i}e_{i}{\rm e}^{-\beta e_{i}\Psi}
\label{eq:PB}
\end{equation}

Alternatively, the first two relations, obtained from the
extremum condition with respect to $\Psi$ and $c_{\rm i}$, can be
substituted into Eq.~(\ref{eq:OmegaPB}). Formally, this gives
$\Omega_{\rm PB}$ as a functional of $\lambda$. Using
Eq.~(\ref{eq:min1}), the expression obtained for $\Omega_{\rm PB}$ 
can be written as a functional of $\Psi$:

\begin{eqnarray}
\Omega_{\rm PB} & = &
   - \frac{\varepsilon}{8\pi} \int_{V} ({\bf \nabla} \Psi)^{2}
         {\rm d}^{3}{\bf r}
   + \oint_{\partial V} \sigma \Psi {\rm d}^{2}{\bf r}_{\rm s}
\nonumber \\
   & & - k_{\rm B}T\int_{V} {\rm d}^{3}{\bf r} \sum_{i}
      \zeta_{i}{\rm e}^{-\beta e_{i} \Psi}
\label{eq:FPBPsi}
\end{eqnarray}

\noindent where the second integration is over the charged 
surfaces. Requiring an extremum of this functional 
with respect to $\Psi$ is another way to obtain the
Poisson-Boltzmann equation~(\ref{eq:PB}).

A more formal derivation of the mean field, PB free energy, 
and a discussion on its
generalization to systems with non-electrostatic interactions
is presented in Ref.~\cite{LZB99}. The PB free
energy (\ref{eq:OmegaPB}) can also be derived by formulating
the problem using field
theory methods. In this approach the mean-field approximation is
obtained as the saddle point of the functional integral, and
corrections due to ion-ion correlations can be obtained in a
systematic expansion \cite{NetzOrland99EPL,NetzOrland00}.

\subsubsection{Inclusion of the hydration interaction}

As discussed in the introduction, our starting point is a model in
which the hydration interaction, arising from solvent effects, is
described as an effective 
ion-pair interaction. We denote this short-range
potential between ions of species $i$ and $j$ at distance {\bf r}
as $u_{ij}({\bf r})$. The potential is taken as the
short-range effective potential between ions immersed in a 
{\it bulk}
ionic solution having a specific, constant concentration.
Therefore, $u_{ij}(\bf r)$ is assumed to be isotropic and
does not depend on the ion positions or the confining geometry.

Our aim is to treat the long range electrostatic interaction on
the mean-field level, as in PB theory. Thus, we begin by
considering the free energy of a system placed in some 
{\it arbitrary} field $\Psi({\bf r})$, where the ions interact
with each other only through the two-body potential $u_{ij}({\bf
r})$. Due to the short-range nature of the hydration interaction,
the free energy can be obtained from a virial expansion of the
grand canonical partition function. Since we will be interested in
highly inhomogeneous systems, we perform an expansion in the
inhomogeneous ion density. The derivation is given in
Appendix~\ref{app:virial}.
Including terms up to the quadratic order in the expansion
we obtain:

\begin{eqnarray}
\Omega_{\rm h} & = & k_{\rm B} T \int \sum_{i} c_{i} \left[ \log
\frac{c_{i}}{\zeta_{i}} - 1 \right] {\rm d}^{3}{\bf r}
\nonumber \\ 
& & + \int
\sum_{i}e_{i} c_{i} \Psi\,{\rm d}^{3}{\bf r} \nonumber \\ & & +
\frac{k_{B}T}{2} \sum_{i,j} \int c_{i}({\bf r})
              U_{ij}({\bf r}-{\bf r'})c_{j}({\bf r'})\,
              {\rm d}^{3}{\bf r}{\rm d}^{3}{\bf r'}
\label{eq:Omegah}
\end{eqnarray}

\noindent where $\Psi({\bf r})$ is an external field, coupled
to the $i$th ion charge density 
$e_{i}c_{i}$. The short-range weighted potential
$U_{ij}$ in the third term of $\Omega_{\rm h}$ is
defined as:

\begin{equation}
U_{ij} = 1 - {\rm e}^{-\beta u_{ij}(|{\bf r}-{\bf r'}|)}
\label{eq:ueff}
\end{equation}

\noindent where $u_{ij}$ is the nominal short-range  
interaction potential
between ions of species $i$ and $j$.
This form of describing the short-range interaction is a rather
crude approximation, valid only in the low density limit. Its
advantage is its simplicity. The free energy $\Omega_{\rm h}$
amounts to setting the direct correlation function $c_{2}(|{\bf
r}-{\bf r'}|)$ to be equal to $-U(|{\bf r}-{\bf r'}|)$,
and all higher order direct correlation functions to zero
\cite{Percus86}.

Having found the hydration free energy $\Omega_{\rm h}$, the
electrostatic interaction can be treated on the mean-field level. 
This is done by considering $\Psi({\bf r})$ as the electrostatic
potential and imposing the self-consistency requirement of the
Poisson equation (\ref{eq:Poisson}). This is essentially the
approximation we used to derive the PB equation (\ref{eq:PB}),
with the difference that the free energy of a dilute,
non-interacting ion distribution is replaced by the free energy
$\Omega_{\rm h}$ of Eq.~(\ref{eq:Omegah}). The result is
the free energy of Eq.~(\ref{eq:Omega}), with $\Delta \Omega$
defined as follows:

\begin{equation}
\Delta \Omega = \frac{k_{B}T}{2} \sum_{i,j} \int c_{i}({\bf r})
U_{ij}({\bf r}-{\bf r'})c_{j}({\bf r'})\,
              {\rm d}^{3}{\bf r}{\rm d}^{3}{\bf r'}
\label{eq:deltaOmega}
\end{equation}

We conclude this section with some remarks on the approach
presented above. Important solvent effects are already introduced in
the PB theory by using an electrostatic interaction with a
dielectric constant $\varepsilon = 78$ of water,
instead of the bare electrostatic interaction. In the modified
model a more precise effective potential between the ions is used.
The separation of this potential into a long-range electrostatic
term and a short-range hydration term allows each of these two
interactions to be treated in a simple though approximated form.
The virial expansion is a standard choice for approximating
short-range interactions. Such an expansion fails for the
electrostatic interaction due to its long-range
\cite{LandauLifshitzSP1}. On the other hand, the wide success of
PB theory demonstrates that the electrostatic interaction can be
treated quite well in the mean-field approximation. Therefore we
use this approximation for the long-range part of the interaction,
and in this respect we remain within the framework of PB theory.

The free energy (\ref{eq:Omega}) can also be obtained by rewriting
the grand canonical partition function as a field-theory partition
function. The short-range part of the interaction can be separated
from the electrostatic interaction and a different expansion can
be performed for each of these interactions. By using a density
expansion for the short-range interaction and a loop expansion for
the electrostatic interaction, Eq.~(\ref{eq:Omega}) is obtained up
to second order in the density expansion and first order in the
electrostatic potential \cite{Netz99pr}.

The simplicity of our approach can lead to elegant analytical
results, but has several limitations. The use of only the second
term in the virial expansion implies that we are using a low
density approximation. The validity of such an approximation for a
bulk fluid can be assessed by considering $k_{\rm B}TB_{2}c$,
where $B_{2}$
is the second virial coefficient in the expansion of 
the pressure, and $c$ is the ion density. Qualitatively, if
$B_{2}c$ is small compared to unity, the correction to the ideal
gas behavior is small and truncating the virial expansion after
the second term is sensible. For non-homogeneous cases, the
corresponding quantity is $(1/2)\sum_{j}\int {\rm d}{\bf
r}'\,c({\bf r}')U_{ij}({\bf r} - {\bf r}')$. For
relatively high surface charges considered here this integral
approaches values of order unity near the charged surfaces,
indicating that the approximation should only be expected to give
qualitative results. Another deficiency of the virial expansion 
to second order can be
seen from the fact that the direct correlation function is simply
$-U_{ij}({\bf r})$. This implies that the hard core
interaction is not described accurately in our treatment. A
faithful description would require the vanishing of the pair
correlation function $h_{2}({\bf r})$ for separations smaller than
the hard-core diameter. Hence, it should be kept in mind that our
main concern is to study the effects of a short-range interaction
with a dominant attractive part. Finally, the fact that we
describe the electrostatic interaction in the mean-field
approximation implies that ion-ion correlations are ignored, as
they are in PB theory. When our approach is applied for the
aqueous pair potential model, these approximations should also be
kept in mind. In particular, we follow
Ref.~\cite{Marcelja97Nature} and do not include an effective
ion-surface potential \cite{TamashiroPincus99}.

\subsection{Density equations}

The mean density distribution is obtained by minimizing the total
free energy $\Omega = \Omega_{\rm PB} + \Delta
\Omega$. From equations (\ref{eq:Omega}), (\ref{eq:OmegaPB}) and
(\ref{eq:deltaOmega}) we have:

\begin{eqnarray}
\Omega & = & \frac{\varepsilon}{8\pi} \int ({\bf \nabla} \Psi)^{2}
   {\rm d}^{3}{\bf r} + k_{B}T
   \int \sum_{i} c_{i} \left( \log
          \frac{c_{i}}{\zeta_{i}} - 1 \right)
   {\rm d}^{3}{\bf r}
\nonumber \\
 & & + \frac{k_{B}T}{2} \sum_{i,j} \int c_{i}({\bf
         r})U_{ij}({\bf r}-{\bf r'})
              c_{j}({\bf r'})\,
              {\rm d}^{3}{\bf r}{\rm d}^{3}{\bf r'}
\nonumber \\
 & & + \int \lambda ({\bf r}) \left(
   {\bf \nabla}^{2}\Psi + \frac{4\pi}{\varepsilon}\sum_{i}c_{i}e_{i}
                              \right)
   \,{\rm d}^{3}{\bf r}
\label{eq:fullOmega}
\end{eqnarray}

\noindent where $\mu_{i}$ and $\zeta_{i} =
\exp(\beta\mu_{i})/\lambda_{\rm T}^{3}$ are the chemical potential
and the fugacity of the ion species $i$, respectively. The
thermal de Broglie wavelength, $\lambda_{\rm T}$, is equal to
$h/(2 \pi m k_{\rm B}T)^{1/2}$, where $h$ is the Planck constant
and $m$ is the ion mass. Requiring an extremum of $\Omega$
with respect to $\Psi$ gives: $\lambda
= (\varepsilon/4\pi)\Psi$ as in Eq.~(\ref{eq:min1}). 
Taking the variation with respect to $c_{i}$
then gives:

\begin{equation}
\log \frac{c_{i}({\bf r})}{\zeta_{i}} + \sum_{j}\int
      c_{j}({\bf r'}) U_{ij}({\bf r}-{\bf r'})\,
      {\rm d}^{3}{\bf r'}
      + \beta e_{i} \Psi ({\bf r}) = 0
\label{eq:c3d}
\end{equation}

\noindent This equation is supplemented by the Poisson
equation~(\ref{eq:Poisson}). Since Eq.~(\ref{eq:c3d}) is an
integral equation, the $c_{i}$ cannot be written as a simple
function of $\Psi$ as in the PB case. Therefore, a single
equation for $\Psi$, analogous to the PB equation, cannot be
obtained, and we are left with the two coupled integro-differential
equations
(\ref{eq:c3d}) and (\ref{eq:Poisson}). These equations should be
solved together to obtain the electrostatic potential and density
profiles. In the case $U \rightarrow 0$,
Eq.~(\ref{eq:c3d}) reduces to the Boltzmann relation
$c_{i}=\zeta_{i}\exp(-\beta e_{i} \Psi)$ with $\zeta_i=c_{{\rm b},i}$. 
Combining this relation
with Eq.~(\ref{eq:Poisson}) reproduces the PB equation
(\ref{eq:PB}).

In order to simplify the set of equations, we assume the same
short-range interaction between the different pairs of ion
species. Assuming that the charged surfaces are negatively
charged, we choose:
$u_{ij}({\bf r})=u_{++}({\bf r}) \equiv u({\bf r})$,
where $u_{++}({\bf r})$ is the short-range effective potential
between the (positive) counter-ions. This assumption is not exact
for the effective potentials of ions in water \cite{Lyubartsev97}.
However, since only the counter-ions reach high densities, close
to the oppositely charged surfaces, and the co-ions are repelled
from the surface neighborhood, the exact choice of the potentials
$u_{+-}({\bf r})$ and $u_{--}({\bf r})$ is expected to be of only
minor significance.

We now consider an electrolyte of valency $z_{+}$:$z_{-}$,
{\it i.e.}, a solution of positive and negative
ions of charges $e_{\pm}=\pm z_{\pm}e$,
where $e$ is the electron charge. We designate the
surface charge density on the plate as a constant 
$\sigma$ and the bulk
densities of the positive and negative ions as $c_{\rm b} \equiv
c_{{\rm b},+}$ and $c_{{\rm b},-}$, respectively.
Due to charge neutrality in
the bulk, $c_{{\rm b},-}=(z_{+}/z_{-})c_{\rm b}$
and similarly, $\zeta_{-} =
(z_{+}/z_{-})\zeta$ where $\zeta \equiv \zeta_{+}$.
Equation~(\ref{eq:c3d}) can then be written as follows:

\begin{equation}
\label{eq:c3d_b} c_{\pm}({\bf r}) =
      \zeta_{\pm} e^{\mp \beta e z_{\pm} \Psi}\exp
      \left[-\int c({\bf r'})U({\bf r}-{\bf r'})
      \,{\rm d}^{2}{\bf r'}\right]
\end{equation}

\noindent where $c({\bf r}) = c_{+}({\bf r}) + c_{-}({\bf r})$
is the total ion density, and $U({\bf r}) = U_{++}({\bf r})$
is obtained from $u({\bf r})$ using Eq.~(\ref{eq:ueff}).
From the Poisson equation (\ref{eq:Poisson}) we obtain:

\begin{eqnarray}
{\bf \nabla}^{2}\Psi & = &
     -\frac{4\pi e}{\varepsilon}
    \left(z_{+}c_{+} - z_{-}c_{-}\right)
    \nonumber \\
    & = & \frac{4\pi e}{\varepsilon} \zeta z_{+}
    \left({\rm e}^{\beta e z_{-} \Psi}-
          {\rm e}^{-\beta e z_{+}\Psi} \right)
    \nonumber \\ & & \ \ \ \times
         \exp \left[-\int c({\bf r'})U({\bf r}-{\bf r'})
         \,{\rm d}^{3}{\bf r'}\right]
\label{eq:Poisson3d}
\end{eqnarray}

\noindent Note that in addition to the explicit dependence on the
ion valencies $z_{\pm}$ in equations~(\ref{eq:c3d_b}) and
 (\ref{eq:Poisson3d}), in a more realistic model
the details of the potential $u({\bf r})$
should also depend on the type of counter-ion species
present in the problem.

\section{Single charged plate}
\label{sec:oneplate}

\subsection{Density equations}

After presenting the general formalism let us consider, as an
example, a single negatively charged planar surface (Fig.~1).
The charged surface is in contact with an electrolyte of valency
$z_{+}$:$z_{-}$. We designate the axis perpendicular to the plate as the
$z$ axis, and consider the ion solution in the region $z
> 0$. For simplicity we consider positive and negative ions of
the same hard-core diameter $d_{\rm hc}$. The coordinate of
closest approach of the ions to the plate is designated as $z =
0$. Hence the ``real'' surface lies at a distance of one ion
radius $d_{\rm hc}/2$ from the actual $z = 0$ plate position, as
shown in Fig.~1a. When we refer to conventional PB results,
however, the ions are point-like and the plate should be
understood to be positioned exactly at $z = 0$.

Due to the one-dimensional symmetry imposed by the uniformly 
charged planar plate, the integration in Eq.\ (\ref{eq:c3d_b}) can be
performed over the $x-y$ plane, leaving us with profiles depending
only on $z$, the distance from the plate:

\begin{equation}
\label{eq:c1d} c_{\pm}(z) =
      \zeta_{\pm} e^{\mp \beta e z_{\pm} \Psi}\exp
      \left[-\int_{0}^{\infty} c(z')B(z-z')\,{\rm d}z'\right]
\end{equation}

\noindent where $c(z) = c_{+}(z)+c_{-}(z)$ is the total ion density and
$B(z)$ is the integral of $U({\bf r})$ in the plane of constant $z$.
Using cylindrical coordinates:

\begin{equation}
B(z)=2\pi \int_{0}^{\infty}\rho\,{\rm d}\rho\,U
                     \left(\sqrt{z^{2}+\rho^{2}}\right)
\label{eq:Bdefinition}
\end{equation}

\noindent
and the Poisson equation (\ref{eq:Poisson3d}) reads:

\begin{eqnarray}
\frac{{\rm d}^{2}\Psi}{{\rm d}z^{2}} & = &
    \frac{4\pi e}{\varepsilon} \zeta z_{+}
    \left({\rm e}^{\beta e z_{-} \Psi}-
          {\rm e}^{-\beta e z_{+}\Psi} \right)
    \nonumber \\ & & \ \ \ \times
         \exp \left[-\int_{0}^{\infty} c(z')B(z-z')\,{\rm d}z'\right]
\label{eq:Poisson1d}
\end{eqnarray}

\noindent Equations (\ref{eq:c1d}) and (\ref{eq:Poisson1d}) are
supplemented by the boundary conditions:

\begin{equation}
\left.\frac{{\rm d}\Psi}{{\rm d}z}\right|_{z=0}=
        -\frac{4\pi}{\varepsilon} \sigma \ \ ; \ \
\left.\frac{{\rm d}\Psi}{{\rm d}z}\right|_{z \rightarrow \infty}=0
\label{eq:boundary1d}
\end{equation}

Finally, the relation between $\zeta$ and the bulk density
$c_{\rm b}$ can be obtained from Eq.~(\ref{eq:c1d}).
As $z \rightarrow \infty$, $\Psi$ becomes zero, and $c_{\pm}$
assume their asymptotic constant, bulk values. Thus the
integrand inside the exponential can be replaced by
$-(1 + z_{+}/z_{-})c_{\rm b}B(z-z')$.
Recalling that $c_{+}=c_{\rm b}$ 
and $c_{-}=(z_{+}/z_{-})c_{\rm b}$, we obtain:

\begin{equation}
c_{\rm b} = \zeta\exp\left[- \left(1+\frac{z_{+}}{z_{-}}\right)
B_{\rm t}c_{\rm b}\right] \label{eq:cb}
\end{equation}

\noindent where:

\begin{equation}
B_{\rm t} \equiv \int_{-\infty}^{\infty}{\rm d}z \, B(z) =
      \int {\rm d}^{3}{\bf r} \, U({\bf r})
\label{eq:btotal}
\end{equation}

\noindent is also equal to $2B_{2}$, the second virial
coefficient.  Note that $B(z)$ and $B_{\rm
t}$ are negative for an attractive interaction. The limit $B_{\rm
t}c_{\rm b} \rightarrow 0$ is the limit in which the short-range
interaction becomes negligible in the bulk. In this limit the
relation between the bulk density and fugacity of
Eq.~(\ref{eq:cb}) tends to the ideal gas relation $c_{\rm b} = \zeta =
\exp(\beta \mu) / \lambda_{\rm T}^3$.

Two special cases will be of particular interest in the following
sections. The first is the case of a monovalent 1:1 electrolyte, where
we have:

\begin{eqnarray}
c_{\pm}(z) & =
    & \zeta {\rm e}^{\mp \beta e \Psi}\exp
      \left[-\int_{0}^{\infty} c(z')B(z-z')\,{\rm d}z'\right]
\nonumber \\ \frac{{\rm d}^{2}\Psi}{{\rm d}z^{2}} & =
    & = -\frac{4\pi}{\varepsilon}c(z)
\label{eq:symmetric}
\end{eqnarray}

\noindent and:

\begin{equation}
c_{\rm b} = \zeta\exp\left(-2 B_{\rm t}c_{\rm b}\right)
\label{eq:cbsymmetric}
\end{equation}

\noindent
The second case is that of no added salt.
The solution contains only monovalent counter-ions
($z_{+} = 1$, $z_{-} = 0$). This case can be obtained by taking
formally
the limit $\zeta \rightarrow 0$ of Eq.~(\ref{eq:symmetric}), or by
repeating the derivation from Eq.~(\ref{eq:fullOmega}) with only
one type of ions, of charge $e$. The term
$-k_{\rm B}T\int{\rm d}^{3}{\rm r}\, c \log (\zeta)$
in $\Omega$ is then a Lagrange multiplier added to impose
the condition: $\int_{0}^{\infty}{\rm d}z\, e c(z) = |\sigma|$. The
following equations are then obtained:

\begin{eqnarray}
c(z) & = & \zeta_{0} {\rm e}^{- \beta e \Psi}\exp
         \left[-\int_{0}^{\infty} c(z')B(z-z')\,{\rm d}z'\right]
\nonumber \\ \frac{{\rm d}^{2}\Psi}{{\rm d}z^{2}} &
    = & -\frac{4\pi e}{\varepsilon}c(z) 
\label{eq:nosalt}
\end{eqnarray}

\noindent where $\zeta_{0}$ is an arbitrary reference fugacity.
The choice of $\zeta_{0}$ determines the (arbitrary) position in
which $\Psi$ is zero. Note that the electrostatic potential $\Psi$
diverges in the bulk. 
This divergence exists also in the usual
PB theory, because the system is effectively one
dimensional with no screening by added salt. Although
$\Psi(z)$ has a weak logarithmic divergence, the density of
counter-ions decays to zero, $\lim_{z \rightarrow \infty}c(z)
=0$
as it should.

\subsection{Parameters and length scales}

For the ion-ion potential $u({\bf r}-{\bf r'})$ we use an
effective potential between \mbox{${\rm Na}^{+}$ - ${\rm Na}^{+}$}
ion pairs. The potential was calculated using a Monte-Carlo simulation
\cite{Lyubartsev97} for an NaCl ionic solution of
concentration $0.55$\,M, at room temperature. The electrostatic
interaction between the ions is subtracted,
and the net short-range potential is
shown in Fig.~2. For ion-ion separations below $2.9$\,{\AA} a hard
core interaction is assumed. Fig.~3 shows the function $B(z)$,
derived from this potential, using Eq.~(\ref{eq:Bdefinition}).
Note that $B(z)$ has several local maxima and minima. These
correspond to the local maxima and minima of $u({\bf r})$.
Thus the structure of $B(z)$ reflects the oscillatory
behavior of the effective potential.

\begin{figure}[tbh]
\epsfxsize=0.9\linewidth \centerline{\hbox{
\epsffile{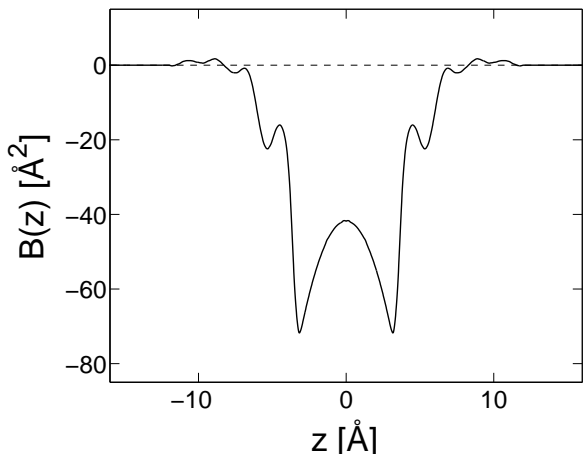} } }
\caption{The effective interaction 
in a planar geometry $B(z)$ obtained
from the potential of Fig.~2, using Eq.~(\ref{eq:Bdefinition}).
The oscillating structure of the radial potential shown in 
Fig.~2 is apparent in the secondary minima of $B(z)$.}
\end{figure}

We use the effective potential calculated for $c_{\rm b} = 0.55$\,M, 
regardless of the actual bulk ion concentration
in the system. Since the important effects occur near the charged
surface, where the ion concentration is much larger than $c_{\rm b}$,
it seems reasonable to use an effective potential calculated
in the presence of a rather high salt concentration. The choice of
$c_{\rm b} = 0.55$\,M 
is still somewhat arbitrary, and we rely on the fact that
the dependence of the effective potential on the ion concentration
is weak \cite{Lyubartsev97}.

It is useful to employ two length scales that characterize
the PB density profiles \cite{Andelman95}. The {\it Gouy-Chapman
length}, defined as $b = \varepsilon k_{\rm B}T/(2\pi e
|\sigma|$), characterizes the width of the diffusive
counter-ion layer close to a single plate charged with a surface
charge $\sigma$, in the absence of added salt. The
{\it Debye-H\"uckel screening length}, $\lambda_{\rm D} = (8\pi
c_{\rm b}e^{2}/\varepsilon k_{\rm B}T)^{-1/2}$, 
equal to $19.6\,\mbox{\AA}$ for $c_{\rm b} = 0.025$M at room
temperature
characterizes the decay of the screened electrostatic interaction 
in a solution with added salt. 
The strength of the electrostatic
interaction can also be expressed using
the {\it Bjerrum length}, $l_{\rm B}
= e^{2}/(\varepsilon k_{\rm B}T)$.
This is the distance at which the
electrostatic interaction between two unit charges becomes equal
to the thermal energy. The Bjerrum length is equal to about
$7$\,{\AA} in water at room temperature.

The inclusion of hydration interactions introduces additional
length scales in the system. For the interaction shown in 
Figs.~2 and 3,
the range of the interaction $d_{\rm hyd}$ can be seen to be
approximately $7$\,\AA, over twice the hard core diameter $d_{\rm
hc} = 2.9$\,\AA. The strength of the hydration interaction is
characterized by $B_{\rm t} \simeq -(7.9\,\mbox{\AA})^3$, as is
calculated from Eq.~(\ref{eq:btotal}).

\subsection{Numerical results}
\label{subsec:densitynum}

Equations (\ref{eq:c1d}) and (\ref{eq:Poisson1d}) are a set of
three nonlinear integro-differential equations. We treat them
numerically using an iterative scheme, based on the assumption
that the positive ion density profile is dominated by the
electrostatic interaction. We start with the
analytically known PB profile close to a single charged plate
and calculate iteratively corrections to this profile, as result
from equations (\ref{eq:c1d}) and (\ref{eq:Poisson1d}). For a 1:1
electrolyte we iteratively solve the equation:

\begin{eqnarray}
\frac{{\rm d}^{2}\Psi^{(n)}}{{\rm d}z^{2}} & = &
     \frac{8\pi e}{\varepsilon} \zeta \sinh\left(\beta e \Psi^{(n)} \right)
\nonumber \\ & & \ \ \times
         \exp\left[-\int_{0}^{\infty} c^{(n-1)}(z')B(z-z')\,{\rm d}z'\right]
\label{eq:iter}
\end{eqnarray}

\noindent where $c(z) = c_{+}(z) + c_{-}(z)$ is the total ion density
and the superscript $n$ stands for the $n$th
iteration. For $n > 0$:

\begin{eqnarray}
c^{(n)}_{\pm}(z) & \equiv & \zeta{\rm e}^{\mp \beta e \Psi^{(n)}}
\nonumber \\ & & \ \times
\exp\left[-\int_{0}^{\infty} c^{(n-1)}(z')B(z-z')\,{\rm d}z'\right]
\end{eqnarray}

\noindent and the zeroth order densities $c^{(0)}_{\pm}$
are taken as
the density profiles generated by the PB equation (\ref{eq:PB}).
The boundary conditions (\ref{eq:boundary1d}) are satisfied by the
electrostatic potential $\Psi^{(n)}$ in all the iterations.
Note that using our iterative scheme,
Eq.~(\ref{eq:iter}) is an inhomogeneous differential equation, because
the integral in the exponential is a known function of $z$,
calculated numerically in the $(n-1)$ iteration. A similar
iterative scheme, based on Eq.~(\ref{eq:nosalt}) can be used
when only counter-ions are present in the solution.

\begin{figure}[tbh]
\epsfxsize=0.9\linewidth \centerline{\hbox{
\epsffile{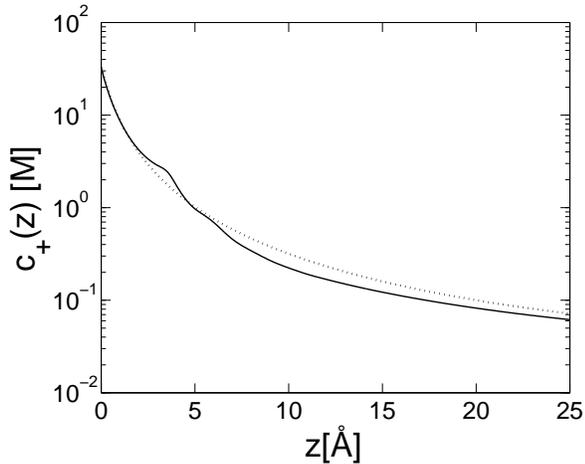} } }
\caption{Counter-ion density profile 
(solid line) obtained from
numerical solution of Eq.~(\ref{eq:nosalt}) with the hydration
interaction as of Fig.~3, plotted on a semi-log plot. No salt is
present in the solution. The surface charge is $|\sigma| = 0.333\,
{\rm C/m^{2}}$. The dielectric constant is $\varepsilon = 78$ and
the temperature is $T = 298$\,K. The dotted line shows the
corresponding density profile obtained from the PB equation.}
\end{figure}

\begin{figure}[tbh]
\epsfxsize=0.9\linewidth \centerline{\hbox{
\epsffile{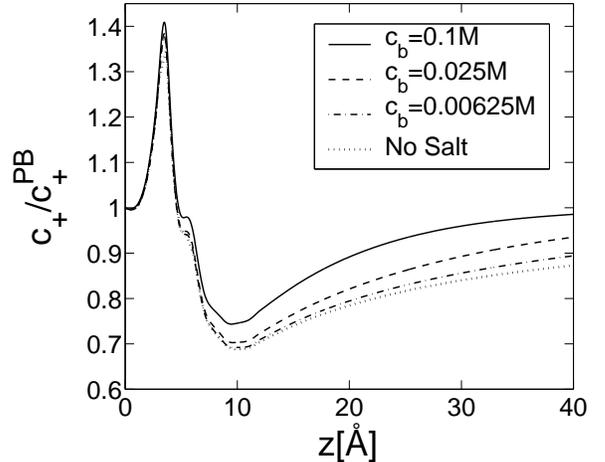} } }
\caption{The ratio $c_+/c_+^{\rm PB}$ between
the positive ion density obtained from Eq.~(\ref{eq:symmetric})
and the value obtained from PB theory, for a surface charge
$|\sigma| = 0.333\,{\rm C/m^{2}}$ and several values of
$c_{\rm b}$. Other parameters
are as in Fig.~4. The three values of $c_{\rm b}$: 
$0.1$\,M, $0.025$\,M and $0.00625$\,M correspond
to Debye-H\"uckel screening lengths $\lambda_{\rm D} \simeq
9.8\,\mbox{\AA}$, $19.6\,\mbox{\AA}$ and $39.2\,\mbox{\AA}$,
respectively.}
\end{figure}

\end{multicols}
\begin{multicols}{2}

Figure~4 shows the calculated density profile of the counter-ions
on a semi-logarithmic scale, for a charged plate with a surface
charge, $|\sigma| = 0.333\,{\rm C/m^{2}}$, corresponding to an area
of approximately $48\,\mbox{\AA}^{2}$ per unit charge. This is a
typical high surface charge obtained with mica plates. It
corresponds to a Gouy-Chapman length $b = 1.06$\,\AA, at a
temperature of $298$K, with $\varepsilon=78$. No salt is present
in the solution. The calculated density profile (solid line) is
compared to the PB prediction (dotted line). The short-range
attraction favors an increased concentration of counter-ions in
the vicinity of the charged plate. This results in an increase of
the concentration relative to the PB prediction. For a surface
charge as in Fig.~4, an increase of the concentration is seen at
distances from the plate up to approximately $4.5\,\mbox{\AA}$. The
overall number of counter-ions is fixed by the requirement of
charge neutrality. Therefore, the increase in the density of
counter-ions {\it near} the plate is balanced by a reduced
concentration {\it further} away.

When salt is present in the solution, the
short-range attraction
draws additional ions from the bulk solution to the diffuse
electrical layer near the plate.
This can be seen in Fig.~5, in a comparison
of counter-ion profiles for different values of the bulk
concentration $c_{\rm b}$.
For each salt concentration, the figure shows the
ratio between the counter-ion density and the density
predicted by PB theory, as a function of the distance
from the plate. The dotted line shows
the result in the no-salt limit.
As the salt concentration increases, the counterion concentration
increases relative to the PB concentration at all distances from
the charged plate. Qualitatively, however,
the hydration effect on the counter-ion profile
is similar in all the curves. As long as the 
Debye-H\"uckel screening length is large
compared to the Gouy-Chapman length, $b = 1.06\,\mbox{\AA}$,
the density profile
in the vicinity of the plate is dominated by the balancing
counter-ions and the salt has only a small effect. 

\begin{figure}[tbh]
\epsfxsize=0.9\linewidth \centerline{\hbox{
\epsffile{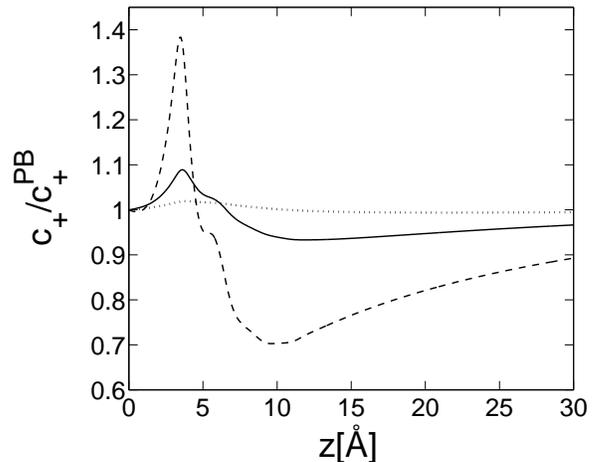} } }
\caption{The ratio $c_+/c_+^{\rm PB}$ between the
positive ion density obtained from Eq.~(\ref{eq:symmetric}) and
the value obtained from PB theory, for surface charges $|\sigma| =
0.333\,{\rm C/m^{2}}$ (dashed line), $0.1\,{\rm C/m^{2}}$ (solid
line) and $0.0333\,{\rm C/m^{2}}$ (dotted line).
The bulk salt concentration $c_{\rm b}$ is $0.025$\,M. Other
parameters are as in Fig.~4.}
\end{figure}

The effect of the hydration interaction is strongly dependent
on the surface charge $\sigma$.
As $\sigma$ is increased, the
ion density near the surface increases too. The exponential in
Eq.~(\ref{eq:nosalt}) deviates more strongly from unity, leading
to a larger deviation from PB theory.
The dependence on $\sigma$ is
 demonstrated in Fig.~6. The ratio of the
positive ion density to its PB value is shown for three values of the
surface charge.
The effect of the hydration potential is very minor for small surface
charge, $|\sigma| = 0.0333\,{\rm C/m^{2}}$ (dotted line), where
the deviation from PB is less than $2\%$ at its maximum, and
considerable for a surface charge of $0.333\,{\rm C/m^{2}}$
(dashed line), where the deviation from PB reaches almost $40\%$. 

The numerical scheme, described above, requires several iterations
to converge fully. It is interesting to note, though, that
the first iteration captures most of the effect of
the short range interaction. This indicates
that the density profile is dominated, as we assumed,
by the electrostatic interaction, and
assures that the convergence of the iterative scheme is good with
the PB density profile as the zero-th order approximation. On the
theoretical level it indicates that the effect of the hydration
interaction can be seen as a perturbation over the PB results.
The fact that the first iteration provides a good approximation
to the full iterative result can lead to further analytical 
approximations. For example, the corrections to the density
profiles, in the no added salt limit, are studied analytically
in the next section, based on this observation.

As an example for the results of the first iteration,
we compare, in Fig.~7, the correction to the counter-ion density
profile obtained in the first iteration (dashed line), 
with the full iterative result (solid line).
We use a high surface charge of
$0.333\,{\rm C/m^{2}}$,  where the differences between the
exact profile and that of the first iteration
are relatively pronounced. The two
density profiles differ by at most $3.2$ percent, where the
ion density deviates from the PB value by $30$ percent.
For smaller surface charge the results obtained in the
first iteration are even better.

\begin{figure}[tbh]
\epsfxsize=0.9\linewidth \centerline{\hbox{
\epsffile{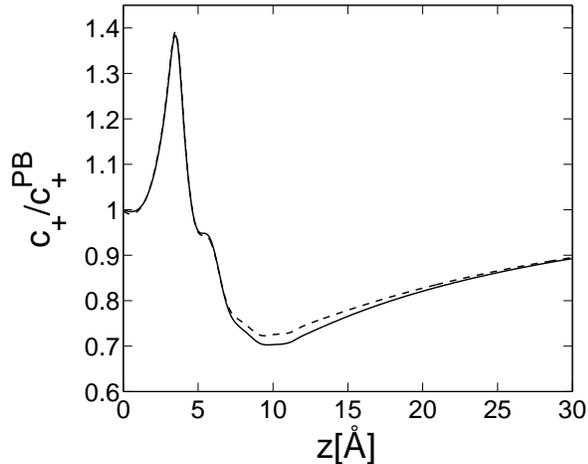} } }
\caption{The positive ion density profile
obtained after one iteration of Eq.~(\ref{eq:iter})
(dotted line), compared to the full solution of
Eq.~(\ref{eq:symmetric}) (solid line). Parameters
are as in Fig.~5. The maximal deviation between the two
density profiles is $3.2$ percent, where the
deviation from PB is approximately $30$ percent.}
\end{figure}

\subsection{Contact density and the contact theorem}

The contact density of the ions is barely modified
as compared with the PB prediction. This is evident in Figs.~4-6.
As long as the Debye-H\"uckel screening length is
large compared to the Gouy-Chapman length, or the
hydration interaction is negligible in the bulk,
the modification remains small. This result
can be obtained from a generalization of the PB contact
theorem \cite{Israelachvili,CarnieChan81}:

\begin{equation}
\sum_{i} c_{i}(0) - \frac{2\pi \beta}{\varepsilon}\sigma^{2}
           = P_{\rm bulk}
\label{eq:contact}
\end{equation}

\noindent where $P_{\rm bulk}$ is the bulk pressure of the
ionic solution. Equation~(\ref{eq:contact}) is derived in
detail for the free energy used in our model
in Ref.~\cite{BurakAndelman00b}. It is obtained from
the equality of the internal pressure in the electrolyte solution
at different distances from the charged plate.
Far away from the charged plate the pressure must be equal to
the bulk pressure of the ionic solution, because the densities
approach their bulk values and the electrostatic potential
becomes constant.
At the contact plane between the plate and the solution,
the pressure involves only an electrostatic contribution and
an osmotic contribution, as in PB theory. This is due to the
fact that in our model no short range interaction between 
the plate and the ions is included. Equating the pressure at the
contact plane and far away from the plate results in
Eq.~(\ref{eq:contact}).

The contact density, as expressed by Eq.~(\ref{eq:contact}),
differs from the PB prediction only due to the change in the
actual value of
$P_{\rm bulk}$. This change is negligible if the
short-range interaction is not of importance in the bulk.
In addition, if the surface
charge is high, such that $b \ll \lambda_{\rm D}$, 
$P_{\rm bulk}$ is negligible compared to the second
term in the left hand side of Eq.~(\ref{eq:contact}).
Thus the contact density remains very close to the PB
prediction. In the no-salt limit $P_{\rm bulk}$ is zero
and the contact density coincides exactly
with the PB result,
$c_{+}(0) = (2\pi \beta / \varepsilon)\sigma^{2}$.

\section{Analytical solutions}
\label{sec:analytical}

The simplicity of the model makes it possible to obtain
various analytical results. The effect of the hydration
on the ion distribution can be characterized by several
quantities, such as the magnitude of the deviation from
the PB result and the effective PB surface charge density
 seen at a distance from the plate. 
Using several simplifying assumptions it is possible 
to obtain analytical expressions for these quantities.

First we assume that the hydration interactions
can be neglected in the bulk, {\it i.e.}, $B_{\rm t}c_{\rm
b} \ll 1$. In this case, the effect of the hydration potential is
significant only in the vicinity of the charged surface, where the
ion density becomes large.
In addition, the
Debye-H\"uckel screening length, $\lambda_{\rm D}$, is taken to
be large
compared to the Gouy-Chapman length $b=e/(2\pi l_{\rm
B}|\sigma|)$. Since $\lambda_{\rm D}
\gg b$, the negative co-ion density near the
negatively charged surface can be neglected compared
to the positive counter-ion density.
Far away from the charged plate,
the system is well described using the PB
equation, with
an effective surface charge density $\sigma_{\rm eff}$ different
from the actual charge density $\sigma$. 
The result of the above two simplifying 
assumptions is
that the salt is of minor importance 
in the region where the effective surface charge
is determined. The
effective surface charge can then be inferred by considering the
case in which only counter-ions are present in the solution
(no added salt).

Equation\ (\ref{eq:nosalt})
can now be recast in a simpler form, by considering
$\eta \equiv \log(c/\zeta_{0})$, as expressed by
Eq.~(\ref{eq:nosalt}),
and taking its second derivative:

\begin{equation}
\frac{{\rm d}^{2}\eta}{{\rm d}z^{2}} =
     \frac{4\pi}{\varepsilon} \beta e^2 \zeta_{0} {\rm e}^{\eta}
             - \int_{0}^{\infty} \zeta_{0}{\rm e}^{\eta(z')}
             \frac{{\rm d}^{2}B(z-z')}{{\rm d}z^{2}}
               \,{\rm d}z'
\label{eq:eta}
\end{equation}

\noindent 
The PB density profile,
$c_{\rm PB}(z) \equiv \zeta_{0} {\rm e}^{\eta_{0}(z)}$,
 for the same surface charge, satisfies
the equation $d^{2}\eta_{0}/dz^{2} = (4\pi \beta e^{2}\zeta_{0} /
\varepsilon)\exp(\eta_{0})$. Its exact solution is known to be:

\begin{equation}
c_{\rm PB}(z) = \zeta_{0}{\rm e}^{\eta_{0}(z)} =
\frac{1}{2\pi l_{\rm B}} \cdot \frac{1}{(z+b)^2} \label{eq:cpb}
\end{equation}

\noindent Note that only in the PB equation
$\eta(z)$ is the reduced electrostatic
potential $e\Psi(z)/k_{\rm B}T$.
From the generalized contact theorem (\ref{eq:contact}),
the surface density in the no added salt case and
in the presence of one plate is
$c(0) = 2\pi\beta\sigma^{2}/\varepsilon$, as in PB theory.
Therefore:

\begin{equation}
\eta(z=0) = \eta_{0}(z=0)
\label{eq:etaboundary1}
\end{equation}

\noindent From the derivative of $c(z)$, Eq.~(\ref{eq:nosalt}), 
we find:

\begin{equation}
\frac{{\rm d}{\eta}}{{\rm d}{z}} = -\beta e\frac{{\rm
d}{\Psi}}{{\rm d}{z}} - \int_{0}^{\infty}{\rm d}z'\,
c(z')\frac{{\rm d}B(z-z')}{{\rm d}z}
\end{equation}

\noindent and using the boundary condition (\ref{eq:boundary1d}):

\begin{equation}
 \left.\frac{{\rm d}\eta}{{\rm d}z}\right|_{z=0} =
  \left.\frac{{\rm d}\eta_{0}}{{\rm d}z}\right|_{z=0}+
  \int_{0}^{\infty}{\rm d}z'\, c(z')\frac{{\rm d}B(z')}{{\rm d}z}
\label{eq:etaboundary2}
\end{equation}

\noindent where the odd parity of ${\rm d}B/{\rm d}z$ has been used.
This relation can be used together with
Eq.~(\ref{eq:etaboundary1}) as a second boundary condition at $z =
0$, instead of the boundary condition of vanishing 
${\rm d}\eta/{\rm d}z$ at infinity.

Linearizing Eq.\ (\ref{eq:eta}) with respect to:

\begin{equation}
w \equiv \eta - \eta_{0} = \log(c/c_{\rm PB})
\end{equation}
 
\noindent which is valid for relatively small deviations from the
PB profile, results in the following equation:

\begin{eqnarray}
& &\frac{{\rm d}^{2}w}{{\rm d}z^{2}}
     -\frac{4\pi}{\varepsilon}\beta e^2 c_{\rm PB}(z)w(z)
\nonumber \\ & &  = - \int_{0}^{\infty}{\rm d}z'\,(1+w(z'))
          c_{\rm PB}(z')\frac{{\rm d}^{2}B(z-z')}{{\rm d}z^{2}}
\end{eqnarray}

\noindent This equation can be further simplified by omitting
$w(z')$ from the integrand in the right hand side. This 
approximation was motivated in Sec.~\ref{subsec:densitynum} 
and is equivalent to stopping the iterative scheme (\ref{eq:iter}) after
the first iteration. The density profile is then replaced by
the PB density profile in the term that involves the hydration
interaction $B(z)$. This results in the equation:

\begin{equation}
\frac{{\rm d}^{2}w}{{\rm d}z^{2}}
     -\frac{4\pi}{\varepsilon}\beta e^2 c_{\rm PB}(z)w(z)
     + \Gamma(z) = 0
\label{eq:linearization}
\end{equation}

\noindent where $\Gamma(z)$ is the convolution integral:

\begin{equation}
\Gamma(z) = \frac{1}{2\pi l_{\rm B}} \int_{0}^{\infty}{\rm d}z'\,
\frac{1}{(z'+b)^{2}} \frac{{\rm d}^{2}B(z-z')}{{\rm d}z^{2}}
\label{eq:conv}
\end{equation}

\noindent The corresponding boundary conditions, obtained from
equations (\ref{eq:etaboundary1}) and (\ref{eq:etaboundary2})
using the same approximations, are:

\begin{eqnarray}
w(z = 0) & = & 0 \nonumber \\ \left.\frac{{\rm d}w}{{\rm
d}z}\right|_{z=0} & = & \int_{0}^{\infty}{\rm d}z'\,c_{\rm
PB}(z')\frac{{\rm d}B(z')}{{\rm d}z} \label{eq:linearboundary}
\end{eqnarray}

\noindent Equation (\ref{eq:linearization}) is a second order
linear differential equation for $w(z)$ and can be solved
analytically. The solution, given in detail in
Appendix~\ref{app:analytical}, is expressed in terms of the
convolution integral $\Gamma(z)$ of Eq.~(\ref{eq:conv}). The
effective surface charge and the effect of the hydration on the
density profile can then be calculated in several limits,
described in detail in Appendix~\ref{app:analytical}. 
Here we outline the main results.

\subsection{Slowly varying density: $b \gg d_{\rm hyd}$}

In the limit $b \gg d_{\rm hyd}$, the PB distribution varies
slowly on the scale of the hydration interaction,
described by $B(z)$, and the theory
becomes effectively a local density functional theory.
The specific form of $B(z)$ is not important,
and all the results simply depend
on $B_{\rm t} = \int_{-\infty}^{\infty}B(z)\,{\rm d}z$.
The deviation of the effective Gouy-Chapman
length $b_{\rm eff}$ from the actual Gouy-Chapman length
$b$ depends linearly on $B_{t}$ and on the surface charge
$\sigma \sim 1/b$. This can be expected since we use
a linearized equation. Thus we have, on dimensional grounds,
$b_{\rm eff} - b \sim B_{t}/l_{B}b$. The detailed
calculation gives the numerical prefactor:

\begin{equation}
b_{\rm eff}-b \cong \frac{-B_{\rm t}}{4\pi l_{\rm B}}\frac{1}{b}
\label{eq:befflargeb}
\end{equation}

\noindent Since $B_{\rm t}$ is negative
$b_{\rm eff}$ is larger than $b$ and
the effective surface charge, $\sigma_{\rm eff}$, is
smaller than the actual surface charge $\sigma$.
This result should be expected. The short range
interaction attracts counterions to the vicinity of the
charged plate and the surface charge is screened more
effectively than in the PB equation.

The correction to the counter-ion density profile,
described by $w(z) = \log[c(z)/c_{\rm PB}(z)]$, is found
to be:

\begin{equation}
w(z) = \frac{-B_{\rm t}}{2\pi l_{\rm B}}
       \left\{\frac{3}{2(z+b)^2} -
       \frac{1}{b(z+b)} \right\}
\label{eq:densitylargeb}
\end{equation}

\noindent The density profile is increased relative to
PB theory for distances smaller than $b/2$, and decreased for
larger distances. The deviation from PB, $w(z)$, is maximal at
$z = 0$, where it is equal to $-B_{\rm t}/(4\pi l_{\rm B}b^{2})$,
and minimal at $z = 2b$, where it is equal to $B_{\rm t}/(12\pi
l_{\rm B}b^{2})$. 

Figure~8 shows the approximated function $w(z)$ of Eq.\
(\ref{eq:densitylargeb}) for $b=21.2$\,{\AA}, corresponding to
$b/d_{\rm hyd} \approx 3$ (dotted line). The approximation is
compared with the function $w(z)$ obtained from the exact solution
of equation (\ref{eq:nosalt}) for the case of no added salt
(solid line). Although $b$ is not much larger than $d_{\rm hyd}$,
the approximation describes well the correction to the PB profile.
Note that $w(z)$, as expressed by
Eq.~(\ref{eq:densitylargeb}) is maximal at $z = 0$, 
whereas according to the contact theorem $w(0)$ should 
be zero. This apparent inconsistency results from neglecting 
the range of the hydration potential relative to $b$. 
In the precise solution of Eq.~(\ref{eq:nosalt}) 
$w(0)$ is zero, as it should be.
The prediction of Eq.~(\ref{eq:densitylargeb}) is valid only 
for distances $z \gtrsim d_{\rm hyd}$, as can be seen in Figure~8.

The range of validity of the linearization procedure
can be found by requiring that the minimal and maximal values of
$w(z)$ are small compared to unity:

\begin{equation}
\frac{-B_{\rm t}}{4\pi l_{\rm B} b^{2}} \ll 1
\end{equation}

\begin{figure}[tbh]
\epsfxsize=0.9\linewidth \centerline{\hbox{
\epsffile{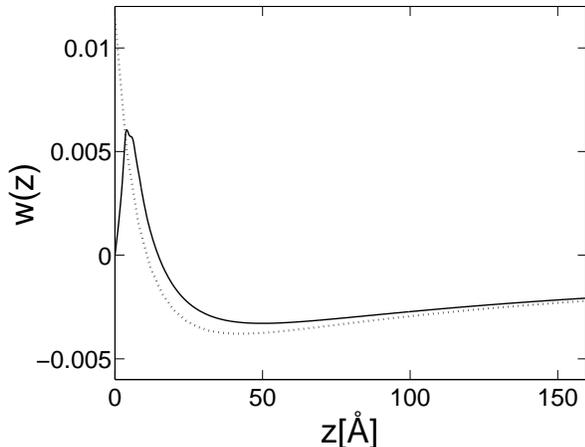} } }
\caption{The logarithm of the ratio between the
counter-ion density obtained with the inclusion of the hydration
interaction and its value in PB theory, $w(z)$, as a function of
the distance from a charged plate, with no added salt in
the solution. The solid line shows the function $w(z)$ obtained
from the exact solution, for $b = 21.2$\,{\AA}.
The dotted line shows the approximated curve
obtained from the linearization with respect to $w$,
Eq.~(\ref{eq:linearization}), in the limit $b \gg d_{\rm hyd}$,
Eq.~(\ref{eq:densitylargeb}).}
\end{figure}

\subsection{Surface layer limit: $b \ll d_{\rm hyd}$}

In the limit in which $b \ll d_{\rm hyd}$, the ion
density effectively becomes a dense layer concentrated 
at $z = 0$ on the
scale of the hydration interaction. The effective Gouy-Chapman
length has the same form as in the limit of slowly varying
density, $b \gg d_{\rm hyd}$, but having a different prefactor:

\begin{equation}
b_{\rm eff}-b \cong \frac{-B_{\rm t}}{12\pi l_{\rm B}}\frac{1}{b}
\label{eq:beffsmallb}
\end{equation}

\noindent The effective surface charge is, therefore, smaller than
the actual surface charge. Note that $b_{\rm eff}$ depends on
$B(z)$, in this limit, only through $B_{\rm t}$. The linear
dependence on $\sigma \sim 1/b$ follows from the linearization
leading to Eq.~(\ref{eq:linearization}), as in the previous limit.

It should be stressed that although $b$ is small compared to
$d_{\rm hyd}$ we still assume that $b$ is large enough for the
linearization to be valid,
{\it i.e.}, we assume that $w(z)$ is small compared to unity.
Furthermore, the counter-ion density should be small enough that we
can sensibly use only the quadratic term in the virial expansion.
To check the validity of these assumptions, the correction to the
density profile should be considered.

The form of $w(z)$ depends, in the surface layer limit, on the
specific form of $B(z)$. In order to study $w(z)$ analytically, we
use an approximated form of $B(z)$, described in
Appendix~\ref{app:analytical}. A typical form of the approximated
$w(z)$, obtained using this approximation
[Eq.~(\ref{eq:densitysmallb})], is shown in Fig.~9 (dotted line).
The Gouy-Chapman length is $b=1.06$\,\AA, corresponding to
$b/d_{\rm hyd} \approx 0.15$. In addition, the function $w(z)$
obtained from the exact solution of equation (\ref{eq:nosalt}) is
shown for comparison (solid line). The approximated curve captures
well the qualitative behavior of the correction to the PB
profile. Note that the discrepancy between the approximated and
actual profiles results not only from the linearization 
and small $b$ limit, but also from the loss of detail 
due to the use of an approximated form for $B(z)$.

The deviation from the PB profile, $w(z)$, can be qualitatively
described as follows. For $z < d_{\rm hc}$, $w(z)$ increases from
zero quadratically (with an additional term of the form $z^{2}\log
z$) to its value at $z = d_{\rm hc}$. It then decreases from its
maximum positive value to a minimum, negative value, on a scale of
the range of the attractive part of $B(z)$. This minimum value is
equal to approximately $B_{\rm t}/6\pi l_{\rm B}b d_{\rm hyd}$.
For distances larger than the interaction range, $w(z)$ assumes
the form $w(z) \sim 1/z$, characterizing a PB profile with a
modified, effective surface charge.  For finite values of $b$, we
can expect the above behavior to be smoothed over a scale of order
$b$.

\begin{figure}[tbh]
\epsfxsize=0.9\linewidth \centerline{\hbox{
\epsffile{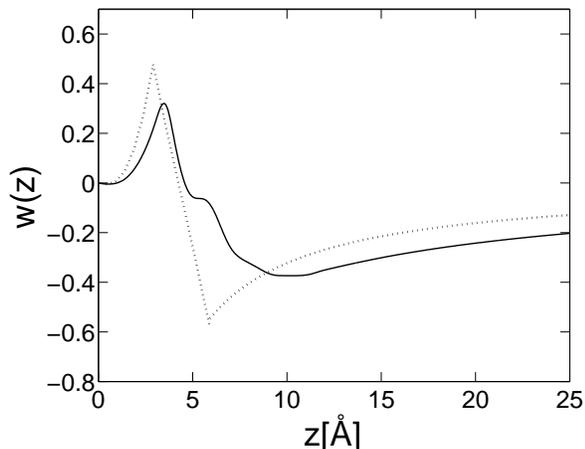} } }
\caption{The logarithm of the ratio between the
counter-ion density obtained with the inclusion of the hydration
interaction and its value in PB theory, $w(z)$, as a function of
the distance from a charged plate, with no added salt in
the solution. The solid line shows the function $w(z)$ obtained
from the exact solution, for $b = 1.06$\,{\AA}. The dotted line 
shows the approximated curve obtained from the linearization 
with respect to $w$, Eq.~(\ref{eq:linearization}), 
in the limit $b \ll d_{\rm hyd}$, Eq.~(\ref{eq:densitysmallb}).
}
\end{figure}

The validity of the linearization can be found by requiring that
$|w(z)| \ll 1$. This requirement results in the following condition:

\begin{equation}
\frac{-B_{\rm t}}{6\pi l_{\rm B} b d_{\rm hyd}} \ll 1
\label{eq:surface_layer_limit}
\end{equation}

\noindent The validity of stopping the virial expansion
at the quadratic order can be shown to have 
the same condition. For the hydration
potential of Fig.~2, the condition expressed in
Eq.~(\ref{eq:surface_layer_limit}) implies that the various
approximations we use start to break down when $b$ becomes smaller
than approximately $1$\,\AA, or 
$\sigma \gtrsim 0.022\,e/{\mbox \AA}^2$. 
When $b$ is of this order, it is well
below $d_{\rm hyd}$, making the surface layer limit a sensible
approximation.

\subsection{Effective surface charge}

In the two limits described above, the effective Gouy-Chapman
length was found to be of the form $b_{\rm eff}-b \sim
-B_{t}/l_{\rm B}b$, with different prefactors in the two limits.
For intermediate values of $b$, the effective charge depends on
the specific structure of the function $B(z)$. In order to study
this dependence, we use a simple approximated form for $B(z)$,
described in Appendix~\ref{app:analytical}. Using this
approximation, an analytical expression can be obtained for the
effective Gouy-Chapman length for all values of $b$.

Figures 10a and 10b show the predicted $b_{\rm eff}$ and $b_{\rm
eff}-b$, respectively (solid lines) as a function of $b$, together
with the asymptotic limits (\ref{eq:befflargeb}) and
(\ref{eq:beffsmallb}) (dotted lines). As the surface charge
increases from zero (and $b$ decreases from infinity), 
the effective
charge $|\sigma_{\rm eff}|$ increases too (but is always smaller
than the actual surface charge). When $b$ reaches a certain value
$b^{\rm min}$, $b_{\rm eff}$ starts increasing with further
reduction of $b$, {\it i.e.}, the effective charge decreases with
increasing surface charge above $|\sigma|^{\rm max} = e/(2\pi
l_{\rm B}b^{\rm min})$. The value of $b^{\rm min}$ depends on the
structure of the function $B(z)$, but can be estimated to be
between the values predicted by the asymptotic expressions
(\ref{eq:befflargeb}) and (\ref{eq:beffsmallb}). From the
condition ${\rm d}b_{\rm eff}/{\rm d}b|_{b = b^{\rm min}} = 0$ we
find:

\begin{equation}
\displaystyle{\sqrt{\frac{-B_{\rm t}}{12\pi l_{\rm B}}} \, < \,
b^{\rm min}}
                \, < \, \sqrt{\frac{-B_{\rm t}}{4\pi l_{\rm B}}}
\end{equation}

\noindent and:

\begin{equation}
b^{\rm min}_{\rm eff} \simeq 2b^{\rm min}
\end{equation}

\end{multicols}

\begin{figure}[b]
\epsfxsize=0.45\linewidth \centerline{\hbox{
\epsffile{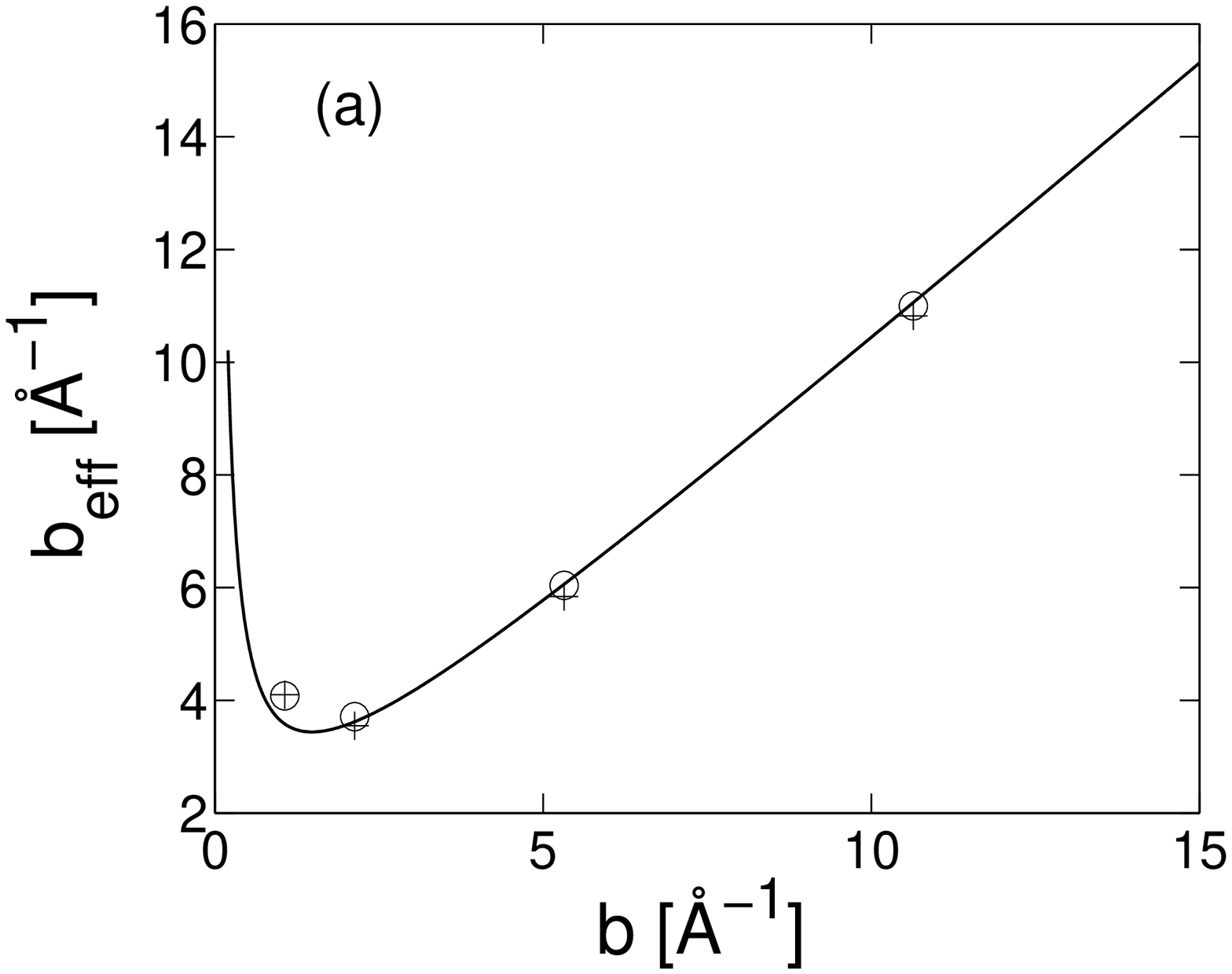} }
\hspace{0.05\linewidth}
\epsfxsize=0.45\linewidth
        \hbox{ \epsffile{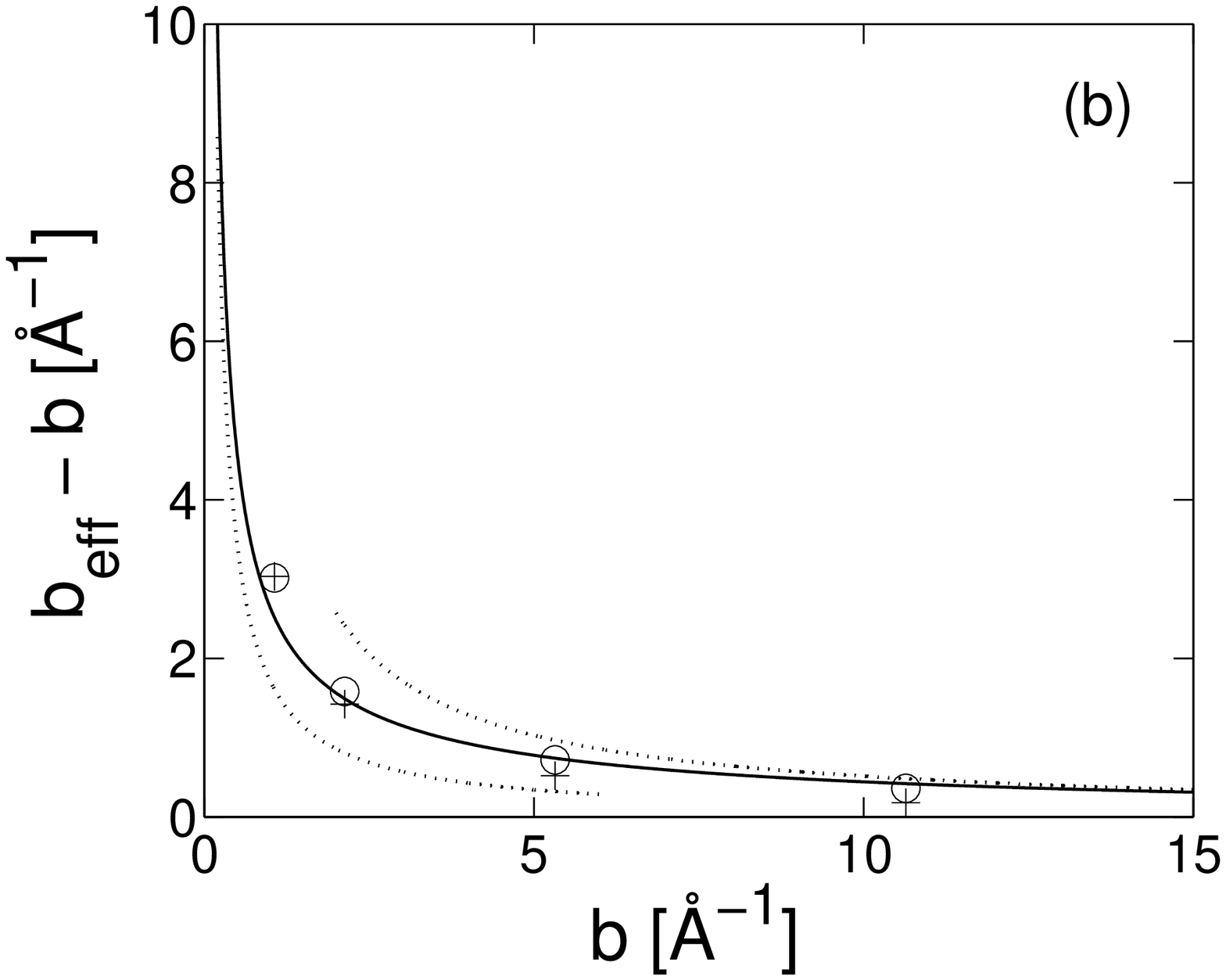} } }
\caption{The effective Gouy-Chapman length $b_{\rm
eff}$ (a) and $\Delta b = b_{\rm eff} - b$ (b), as a function of
the Gouy-Chapman length $b$. The solid lines show the behavior
predicted by Eq.\ (\ref{eq:beffapprox}), with $B_{\rm t} = -500\,
\mbox{\AA}^{3}$, $d_{\rm hc} = 2.9\,\mbox{\AA}$ and $B_{0} = 41.8
\,\mbox{\AA}^{2}$. The dotted lines show the asymptotic limits of
equations (\ref{eq:befflargeb}) and (\ref{eq:beffsmallb}). The
symbols show results extracted from numerical solutions of Eq.\
(\ref{eq:symmetric}), using $B(z)$ of Fig.~3, with
salt concentrations of $10^{-7}$\,M (circles) and $0.1$\,M (crosses).
The salt has a very small effect.}
\end{figure}

\begin{multicols}{2}

\noindent For the hydration interaction of Fig.~2, $B_{\rm t}$ is
approximately $-500\,\mbox{\AA}^{3}$. The value of $b^{\rm min}$
is then between $1.36$\,{\AA} and $2.35$\,\AA, corresponding to a
surface charge density between $0.15\,\rm{C/m^{2}}$ and $0.26\,
\rm{C/m^{2}}$. The values obtained from the approximated curve,
shown in Fig.~10, are $b^{\rm min} \simeq 1.5 \,\mbox{\AA}$ and
$b^{\rm min}_{\rm eff} \simeq 3.4$\,\AA.

For small enough values of $b$, the effective surface charge
$|\sigma_{\rm eff}|$ should increase again with an increase of
$|\sigma|$ and become larger than $|\sigma|$.
This effect cannot be predicted
by our model because of the low density approximation used for the
hard core interaction. In particular, the hard core of the ions 
should cause the density to saturate at the
close packing density, leading
to a reduced screening of the surface charge relative to PB
theory \cite{BAO97,BAO00,Kjellander96}. 
In our model, as in the PB theory, the counterion density 
near the surface is not bounded, and increases indefinitely 
as $\sigma$ is increased. Although our model includes the steric 
repulsion between ions, this
repulsion is ``softened'', and is always outweighed 
by the attractive part of the ion-ion interaction.

In addition to the prediction obtained using the linearized
approximation, \noindent Figure~10 shows values of $b_{\rm eff}$
extracted from numerical solutions of the full equation
(\ref{eq:symmetric}), using the original interaction $B(z)$. The
equation was solved with two different salt concentrations:
$10^{-7}$\,M (circles) and $0.1$\,M (crosses). The value of
$b_{\rm eff}$ was estimated from the positive ion density at large
distances from the plate, by finding the value of $b$ that would
result in the same calculated values of the density in a solution
of the PB equation. Note that for both salt concentrations, 
$b_{\rm eff}$ is very close to its predicted value, meaning that 
the salt has a very small effect on $\sigma_{\rm eff}$.
This result is not obvious for the high salt concentration of 
$0.1\,{\rm M}$. The
Debye-H\"uckel screening length is approximately $9.6$\,\AA, not
much larger than the range of the hydration interaction, 
$d_{\rm hyd} \simeq 7\,\mbox{\AA}$, and
comparable to the Gouy-Chapman length at the large $b$ region of
the plot.

\section{Conclusions and Outlook}

In this work we have studied the effects due to the discreteness
of the solvent in aqueous ionic solutions. 
Hydration interactions are found to have a significant 
effect on the structure
of the diffusive layer near highly charged surfaces. The
counter-ion density is increased in the vicinity of the charged
surface, relative to the PB prediction, and decreased further
away. The distance from the charged plate in which the density
is increased, and the magnitude of the deviation from the PB 
density, depend strongly on the surface charge, and on the
parameters of the short-range hydration interaction
between ion pairs.

The ion-ion hydration interaction can be described roughly using
two parameters. The first parameter is the range of the 
hydration interaction, 
$d_{\rm hyd}$, equal to approximately $7\,\mbox{\AA}$ for 
Na$^+$-Na$^+$ pairs.
The second parameter, $B_{\rm t}$ has dimensions of 
volume and characterizes the strength of the hydration
interaction. It is equal to approximately 
$-500\,\mbox{\AA}^{3}$ for Na$^+$-Na$^+$ pairs. 
Two limits can be considered, where
the Gouy-Chapman length, $b \sim 1/\sigma$,
is small or large compared to the range of the hydration
interaction $d_{\rm hyd}$. In both of these limits we
assume that the Debye-H\"uckel screening length, 
$\lambda_{\rm D}$, is large compared to $b$ and $d_{\rm hyd}$.

In the limit $b \gg d_{\rm hyd}$, 
the counter-ion density becomes depleted,
relative to the PB prediction, 
starting at a distance $z \simeq b/2$ from the
charged plate. The maximum absolute value of 
$w(z) = \log[c(z)/c_{\rm PB}(z)]$ scales as 
$-B_{\rm t}/l_{\rm B}b^{2}$. In the limit $b \ll d_{\rm hyd}$,
the distance from the plate, where the counter-ion density
becomes lower than the PB prediction, is between
$z = d_{\rm hc}$ and $z = d_{\rm hyd}$.
The maximum absolute value of $w(z)$ scales as 
$-B_{\rm t}/l_{\rm B}d_{\rm hyd}b$.

Far away from the charged plate, the density profile can be
well described using the PB theory with an effective surface charge
that can be calculated analytically. The correction
to the Gouy-Chapman length in the two limits $b \gg d_{\rm hyd}$ and
$b \ll d_{\rm hyd}$ is always positive and scales as
$-B_{\rm t}/l_{\rm B}b$, but has different numerical prefactors. 
When the surface charge on the plate is increased, the effective
surface charge, $\sigma_{\rm eff}$,
is found to reach a certain maximal value. Above
this maximal value $\sigma_{\rm eff}$ decreases with
further increase of the actual $\sigma$ on the plate. 
The various approximations we use start to break down when
$b$ is smaller than approximately 
$-B_{\rm t}/6\pi l_{\rm B} d_{\rm hyd}$, corresponding to
$b \lesssim 1\,\mbox{\AA}$.

An important outcome of this work is that the correction
of the PB ion density due to the hydration interaction
 is significant near highly charged surfaces. 
The electrostatic
interaction dominates the ionic distribution and
the hydration interaction can be seen as a perturbation.
For a high surface charge density
of, say, one unit charge per $48\,\mbox{\AA}^{2}$ the counterion
density deviates from its Poisson Boltzmann value by at most 
$30$ percent. The effective change in the surface charge is
more significant, from $1\,e/48\,\mbox{\AA}^{2}$
to about $1\,e/13\,\mbox{\AA}^{2}$. 

The hydration effect on inter-surface forces can be very
pronounced, as opposed to the effect on the ion distribution.
This result will be presented elsewhere \cite{BurakAndelman00b}.
Our model predicts an attractive
contribution to the pressure between two 
parallel charged plates. At distances below several nanometers
this contribution can outweigh the electrostatic
repulsion and lead to an overall attraction between the plates.
Our two-plate findings can also be compared 
with available AHNC results \cite{Marcelja98,Marcelja00pr}, 
showing good qualitative agreement both for the ion density 
profile and pressure.

The formalism
we present can be readily generalized to other geometries.   
This could lead to an estimation of the aqueous 
solvent effects on phenomena such as the Manning condensation
on cylindrical polyions \cite{Manning69},
and charge renormalization of spherical mycelles or colloids
\cite{ACGMPH84}.
In this respect our formalism offers an
advantage over the AHNC approximation which was applied
so far only in a planar geometry. 
Another interesting extension of this work would be to
consider the combination of fluctuation and hydration 
effects. This is particularly important for ionic solutions
with divalent counter-ions, where fluctuation effects
become large \cite{GJWL84,KAJM92,StevensRobbins90}. 

\acknowledgements

We wish to thank S. Mar\u{c}elja for introducing us to the subject
of solvent effects in aqueous ionic solutions, and for
valuable discussions and suggestions. We would like to thank R.
Netz, H. Orland and R. Podgornik for useful discussions. Partial
support from the U.S.-Israel Binational Foundation (B.S.F.) under
grant No. 98-00429, and the Israel Science Foundation founded by
the Israel Academy of Sciences and Humanities - Centers of
Excellence Program is gratefully acknowledged.

\end{multicols}

\appendix

\section{Inhomogeneous virial expansion}
\label{app:virial}

We consider an inhomogeneous system of particles with a
short-range two-body
interaction, and aim to express the free energy of the system
in the low density limit as a functional of the
density distribution.
For simplicity we consider only one species of particles.
The inhomogeneity of the system arises from
the inclusion of an external field $\varphi({\bf r})$,
or from the boundary conditions imposed on the system.
We begin by considering the grand canonical ensemble.
The grand canonical partition function is:

\begin{equation}
Z_{\rm G} = \sum_{N}\frac{1}{N!}
        \left(\frac{{\rm e}^{\beta \mu}}
        {\lambda_{\rm T}^{3}}\right)^{N}Q_{N}
\label{eq:appa_zg}
\end{equation}

\noindent where $\mu$ is the chemical potential,
$\lambda_{\rm T}$ is the de Broglie thermal wavelength
and $Q_{N}$ is:

\begin{equation}
Q_{N} = \int \prod_{i=1}^{N}{\rm d}^{3}{\bf r}_{i}\,
        {\rm e}^{-\beta U_{N}(\{{\bf r}_{i}\})}
\end{equation}

\begin{equation}
U_{N}(\{{\bf r}_{i}\}) =
    \sum_{i} \varphi({\bf r}_{i})
  + \frac{1}{2}\sum_{i}\sum_{j \neq i}
    u\left(\left|{\bf r}_{i}-{\bf r}_{j}\right|\right)
\end{equation}

\noindent We proceed on similar lines as the usual virial
expansion in a bulk fluid, expanding $\log Z_{\rm G}$ in
powers of the activity. Up to second order we have:

\begin{eqnarray}
\log Z_{\rm G} & = & \left(\frac{{\rm e}^{\beta \mu}}
       {\lambda_{\rm T}^{3}}
       \right)Q_{1} + \frac{1}{2}
       \left(\frac{{\rm e}^{\beta \mu}}{\lambda_{\rm T}^{3}}\right)^{2}
       \left( Q_{2}-Q_{1}^{2} \right)
   =   \left(\frac{{\rm e}^{\beta \mu}}{\lambda_{\rm T}^{3}}\right)
       \int {\rm d}^{3}{\bf r}\,{\rm e}^{-\beta \varphi({\bf r})}
\nonumber \\
 & & + \frac{1}{2}
       \left(\frac{{\rm e}^{\beta \mu}}{\lambda_{\rm T}^{3}}\right)^{2}
       \int {\rm d}^{3}{\bf r} \int {\rm d}^{3}{\bf r'} \,
       {\rm e}^{-\beta \left( \varphi({\bf r}) +
              \varphi({\bf r'})\right) }
       \left({\rm e}^{-\beta u\left(\left|{\bf r}-
              {\bf r'}\right|\right)} - 1 \right)
\label{eq:appa_ZGmu}
\end{eqnarray}

\noindent This can be seen as an expansion in powers
of the field
$\exp \left[\beta \left(\mu - \varphi({\bf r})\right)
\right]/\lambda_{\rm T}^{3}$.
The local density $c({\bf r})$ can be expressed in a
similar expansion:

\begin{eqnarray}
c({\bf r}) & = & - \frac{1}{\beta}
     \frac {\delta \log Z_{\rm G}}{\delta \varphi({\bf r})}
  =  \left(\frac{{\rm e}^{\beta \mu}}{\lambda_{\rm T}^{3}}\right)
     {\rm e}^{-\beta \varphi({\bf r})}
\nonumber \\
 & & + \left(\frac{{\rm e}^{\beta \mu}}{\lambda_{\rm T}^{3}}\right)^{2}
    {\rm e}^{-\beta \varphi({\bf r})} \int {\rm d}^{3}{\bf r'}
    {\rm e}^{-\beta \varphi({\bf r'}) }
    \left({\rm e}^{-\beta u\left(\left|{\bf r}-
              {\bf r'}\right|\right)} - 1 \right)
\end{eqnarray}

\noindent This relation can be inverted to obtain an expansion
of $\exp \left[\beta \left(\mu - \varphi({\bf r})\right)
\right]/\lambda_{\rm T}^{3}$ in powers of $c({\bf r})$. Up to
the second order:

\begin{equation}
\frac{{\rm e}^{\beta(\mu - \varphi({\bf r}))}}{\lambda_{\rm T}^{3}} =
  c({\bf r}) + c({\bf r})\int {\rm d}^{3}{\bf r'}\,
  c({\bf r'}) \left(1 - {\rm e}^{-\beta u\left(\left|{\bf r}-
                  {\bf r'}\right|\right)} \right)
\label{eq:appa_expandmu}
\end{equation}

\noindent and by substituting this relation in
Eq.~(\ref{eq:appa_ZGmu}) $\log Z_{\rm G}$ can be expressed as an
expansion in $c$. Up to the second order:

\begin{equation}
\log Z_{\rm G} = \int  {\rm d}^{3}{\bf r}\,c({\bf r}) +
\frac{1}{2}\int {\rm d}^{3}{\bf r}\int {\rm d}^{3}{\bf r'}\,
c({\bf r})c({\bf r'})
    \left(1 - {\rm e}^{-\beta u\left(\left|{\bf r}-
              {\bf r'}\right|\right)} \right)
\label{eq:app_ZGc}
\end{equation}

\noindent The grand canonical potential can be obtained from the
relation $\Omega = -k_{\rm B}T\log Z_{\rm G}$, with $\log Z_{\rm G}$
given by Eq.~(\ref{eq:app_ZGc}). In this expression, $c({\bf r})$
is the mean density profile for the imposed external field
$\varphi({\bf r})$ and a given chemical potential $\mu$. We would
like to express $\Omega$ as a functional of a general ion density
$c({\bf r})$, whose minimization with respect to $c({\bf r})$
would give the equilibrium mean density. Regarding $-k_{\rm B}T\log
Z_{\rm G}$ as a functional of $\chi({\bf r}) \equiv \varphi({\bf
r}) - \mu$, we have:

\begin{equation}
-k_{\rm B}T\frac{\delta \log Z_{\rm G}} {\delta \chi({\bf r})}
= c({\bf r})
\end{equation}

\noindent The Legendre transform of this relation can be obtained
by defining:

\begin{equation}
\Theta = -k_{\rm B}T\log Z_{\rm G} - \int{\rm d}^{3}{\bf r}\,c({\bf
r})\chi({\bf r})
\end{equation}

\noindent and expressing $\log Z_{\rm G}$ and $\chi$ as functionals
of $c({\bf r})$. We have already expressed $\log Z_{\rm G}$ as a
functional of $c({\bf r})$ in Eq.~(\ref{eq:app_ZGc}). An
expression for $\chi({\bf r})$ as a functional of $c({\bf r})$ can
be obtained from Eq.~(\ref{eq:appa_expandmu}). Up to first order
in $c$ we have :

\begin{eqnarray}
\beta[\varphi({\bf r})-\mu] & = & -\log
  \left\{ \lambda_{\rm T}^{3} c({\bf r}) \left[
   1 + \int {\rm d}^{3}{\bf r'}\,
   c({\bf r'}) \left(1-{\rm e}^{-\beta u\left(\left|{\bf r}-
                   {\bf r'}\right|\right)} \right)
  \right] \right\}
\nonumber \\
 & = & - \log \left[
    \lambda_{\rm T}^{3}c({\bf r}) \right] -
    \int {\rm d}^{3}{\bf r'}\, c({\bf r'})\left(
    1 - {\rm e}^{-\beta u\left(\left|{\bf r}-
            {\bf r'}\right|\right)}  \right)
    + {\rm O}(c^{2})
\end{eqnarray}

\noindent Using this relation and Eq.~(\ref{eq:app_ZGc}) we
obtain, up to second order in c:

\begin{eqnarray}
\beta \Theta(\{c({\bf r})\}) & = & \int {\rm d}^{3}{\bf r} \,
    c({\bf r})\left[\log(\lambda_{\rm T}^{3}c({\bf r}))-1\right]
\nonumber \\
 & & + \frac{1}{2}\int {\rm d}^{3}{\bf r}
      \int {\rm d}^{3}{\bf r'} \,
      c({\bf r})c({\bf r'})\left(
      1 - {\rm e}^{-\beta u\left(\left|{\bf r}-
            {\bf r'}\right|\right)}  \right)
\label{eq:theta}
\end{eqnarray}

\noindent The functional $\Theta$ of $c({\bf r})$ has the property
that:

\begin{equation}
\frac{\delta \Theta} {\delta c({\bf r})} = - \chi({\bf r}) =
 -\left[\varphi({\bf r})-\mu \right]
\end{equation} or equivalently:

\begin{equation}
\frac{\delta}{\delta c({\bf r})}\left\{\Theta + \int {\rm
d}^{3}{\bf r}\,c({\bf r})\left[ \varphi({\bf r}) -\mu
\right]\right\} = \frac{\delta \Omega(\{c({\bf r})\})}{\delta
c({\bf r})} = 0
\end{equation}

\noindent Thus, using Eq.~(\ref{eq:theta}), we obtain:

\begin{eqnarray}
\Omega(\{c({\bf r})\}) & = & k_{\rm B}T\int {\rm d}^{3}{\bf r} \,
    c({\bf r})\left(\log \frac{c({\bf r})}{\zeta}-1 \right)
    + \int {\rm d}^{3}{\bf r} \, c({\bf r})\varphi({\bf r})
\nonumber \\
 & & + \frac{1}{2}k_{\rm B}T\int {\rm d}^{3}{\bf r}
      \int {\rm d}^{3}{\bf r'} \,
      c({\bf r})c({\bf r'})\left(
      1 - {\rm e}^{-\beta u\left(\left|{\bf r}-
            {\bf r'}\right|\right)}  \right)
\label{eq:Omegac}
\end{eqnarray}

\noindent where $\zeta = \exp (\beta \mu)/ \lambda_{\rm T}^{3}$.

The derivation of Eq.~(\ref{eq:Omegac}) can be readily generalized
to the case of several ion species of different charges and
different pair interactions $u_{ij}({\bf r})$, resulting in
Eq.~(\ref{eq:Omegah}).

A similar, more elaborate diagrammatic expansion of the
thermodynamic potentials in the presence of an external field is
presented in Ref.~\cite{MorHir61}. A variational principal for the
grand canonical potential $\Omega$ is obtained in which $\Omega$
is expressed as a functional of the mean density $c({\bf r})$ and
the pair correlation function $h_{2}({\bf r}_{1},{\bf r}_{2})$.
This expression is equivalent to Eq.~(\ref{eq:Omegac}) up to the
second order in the cluster expansion.

\newpage

\section{Details of analytical results}
\label{app:analytical}

In this appendix we present details of the analytical
approximations of Sec.~\ref{sec:analytical}.

We consider first the analytical solution of Equation
(\ref{eq:linearization}). This equation is a second order linear
differential equation for $w(z)$. Note that the function $c_{\rm
PB}(z)$ is a known function of $z$, given by Eq.~(\ref{eq:cpb}).
The solution of Eq.~(\ref{eq:linearization}), with the boundary
conditions of Eq.~(\ref{eq:linearboundary}) is:

\begin{equation}
w(z)=\frac{1}{z+b} \int_{0}^{z} {\rm d}z_{2} \,
      \left (z_{2}+b\right)^2 \int_{z_{2}}^{\infty}
      \frac{{\rm d}z_{1}}{\left(z_{1}+b\right)}\, \Gamma(z_{1})
\label{eq:wofz}
\end{equation}

\noindent where $\Gamma(z)$ is the convolution integral, defined
by Eq.~(\ref{eq:conv}). By writing $\Gamma(z)$ as:

\begin{equation}
\Gamma(z) = \int_{0}^{\infty}{\rm d}z'\,\Gamma(z')\delta(z-z')
\end{equation}

\noindent $w(z)$ can be rewritten in the following form:

\begin{eqnarray}
w(z) & = & - \frac{1}{z+b}\left\{
         \frac{b^3}{3}\int_{0}^{\infty}{\rm d}z'\,
               \frac{\Gamma(z')}{z'+b}
       - \frac{1}{3}\int_{0}^{z}{\rm d}z'\,
               (z'+b)^2\Gamma(z') \right\}
\nonumber \\
     & & +\frac{(z+b)^2}{3}\int_{z}^{\infty}{\rm d}z'\,
         \frac{\Gamma(z')}{z'+b}
\label{eq:wofz2}
\end{eqnarray}

\noindent The effective charge $\sigma_{\rm eff}$ (or
equivalently, the effective Gouy-Chapman length $b_{\rm eff})$ can
be calculated from the coefficient of $z^{-1}$ in $w(z)$, as $z$
approaches infinity:

\begin{equation}
w(z) \sim \frac{2\left(b-b_{\rm eff}\right)}{z}
        \; , \ \ \ z \rightarrow \infty
\nonumber
\end{equation}

\noindent We thus find:

\begin{equation}
b_{\rm eff}-b = \frac{1}{6}\int_{0}^{\infty}{\rm d}z \,
                \left[ \frac{b^{3}}{z+b} - (z+b)^2  \right]
                \Gamma(z)
\label{eq:beff}
\end{equation}

\noindent A simple form for the convolution integral $\Gamma(z)$
can be obtained in the limits in which $b$ is small or large
relative to $d_{\rm hyd}$, the characteristic range of the
hydration potential.

\subsection{Slowly varying density: $b \gg d_{\rm hyd}$}

In the limit $b \gg d_{\rm hyd}$, the PB distribution varies
slowly on the scale of the hydration interaction. The convolution
integral $\Gamma(z)$ of Eq.~(\ref{eq:conv}) can then be
approximated in the following way:

\begin{eqnarray}
\Gamma(z) & = & \frac{1}{2\pi l_{\rm B}}
\int_{-\infty}^{\infty}{\rm d}z'\, \frac{H(z')}{(z'+b)^2}
\frac{{\rm d}^{2}B}{{\rm d}z^{2}}(z-z') \nonumber \\ & = &
\frac{1}{2\pi l_{\rm B}} \int_{-\infty}^{\infty}{\rm d}z'\,
\left[ \frac{1}{b^{2}}\frac{{\rm d}\delta(z)}{{\rm d}z}
  - \frac{2}{b^{3}}\delta(z) +
  \frac{6H(z)}{(z+b)^4} \right] B(z-z')
\nonumber \\ & \simeq & \frac{B_{\rm t}}{2\pi l_{\rm B}}
     \cdot \left[
       \frac{1}{b^{2}}\frac{{\rm d}\delta(z)}{{\rm d}z}
     - \frac{2}{b^{3}}\delta(z)
     + \frac{6H(z)}{(z+b)^{4}} \right]
\label{eq:convlargeb}
\end{eqnarray}

\noindent where $H(z)$ is the Heaviside function ($H(z)$ = 0 for
$z < 0$ and H(z) = 1 for $z > 0$). Inserting this expression in
Eq.~(\ref{eq:beff}) we obtain Eq.~(\ref{eq:befflargeb}) for the
effective Gouy-Chapman length. By substituting equation
(\ref{eq:convlargeb}) in Eq.~(\ref{eq:wofz2}), the form of $w(z)$,
given in Eq.~(\ref{eq:densitylargeb}), is obtained.

\subsection{Approximated form for $B(z)$}

Some of the following results depend on the specific structure of
the hydration interaction, characterized by the function $B(z)$.
In order to obtain analytical expressions, we use a simple
approximated form, $B^{\rm app}(z)$, instead of $B(z)$. Assuming
that the hydration interaction consists of a hard core interaction
and a short-range attractive part, the function $B(z)$ has some
general characteristics that should be present in $B^{\rm
app}(z)$. For $z < d_{\rm hc}$, $B(z)$ always has the 
parabolic form
$-(B_{0} + \pi z^{2})$, where $B_{0} = -B(z = 0)$. We assume that
the attractive part of the interaction dominates over the
short-range repulsion so that $B_{0}$ is positive. For $z$ larger
than some finite value $d_{\rm hc} + \Delta$, $B(z)$ is
practically zero due to the short range of the interaction. For
$d_{\rm hc} < z <  d_{\rm hc} + \Delta$, $B(z)$ varies from
$-(B_{0} + \pi d_{\rm hc}^{2})$ to zero in a functional form that
depends on the details of the attractive potential. The most
simple way to model this behavior of $B(z)$ is to have a linear
increase of $B^{\rm app}(z)$ between $z = d_{\rm hc}$ and $z =
d_{\rm hc} + \Delta$, and to set $B^{\rm app}$ to be zero for $z$
larger than $d_{\rm hc} + \Delta$:

\begin{equation}
B^{\rm app}(z)= \left\{ \begin{array}{ll}
                     -(B_{0} + \pi z^{2}) \ & \ \
                                  \left| z \right| \le d_{\rm hc} \\
      -\left(B_{0} + \pi d_{\rm
      hc}^{2}\right)\frac{\displaystyle \left( d_{\rm hc}+\Delta-z \right)}
                {\displaystyle \Delta}\ & \ \
                d_{\rm hc} < \left| z \right| \le d_{\rm hc} + \Delta \\
                0 \        & \ \ d_{\rm hc}+\Delta < \left| z \right|
         \end{array}
     \right.
\label{eq:bapp}
\end{equation}

\noindent The parameters in this expression should be chosen to
match, approximately, the form of $B(z)$. Setting $d_{\rm hc}$ to
be the hard core diameter of the real potential and setting $B_{0}
= -B(0)$ ensures that $B(z)$ and $B^{\rm app}(z)$ are identical
for $z < d_{\rm hc}$. The width $\Delta$ can then be set such that
$B^{\rm app}_{\rm t} = B_{\rm t}$:

\begin{equation}
2B_{0}d_{\rm hc} + \frac{2}{3}d_{\rm hc}^{3} +
\Delta\left(B_{0}+\pi d_{\rm hc}^{2}\right) = -B_{\rm t}
\end{equation}

\noindent This is desirable in light of equations
(\ref{eq:befflargeb}) and (\ref{eq:beffsmallb}),
since the effective surface charge depends only on
$B_{\rm t}$ in these limits. Figure~11 shows
$B(z)$ and $B^{\rm app}(z)$ for the hydration potential of Fig.~2.

\begin{figure}[tbh]
\epsfxsize=0.45\linewidth \centerline{\hbox{
\epsffile{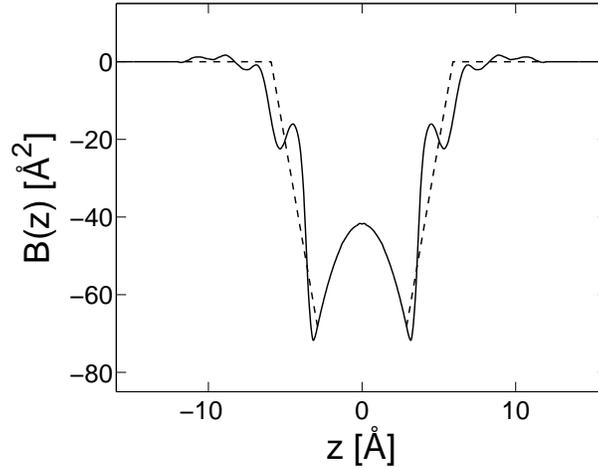} } }
\caption{The effective interaction in a planar
geometry, $B(z)$, obtained
from the potential of Fig.~2, and the corresponding 
approximated function $B^{\rm app}(z)$, defined by 
Eq.~(\ref{eq:bapp}) (dashed line). The parabolic dependance
for $|z| < d_{\rm hc}$ is identical in the two curves.}
\end{figure}

\subsection{Surface layer limit: $b \ll d_{\rm hyd}$}

In the limit $b \ll d_{\rm hyd}$, the convolution
integral in Eq.~(\ref{eq:conv}) becomes:

\begin{equation}
\Gamma(z) \simeq \frac{|\sigma|}{e}
            \frac{{\rm d}^{2}B(z)}{{\rm d}z^{2}}
    = \frac{1}{2\pi l_{\rm B}b}\frac{{\rm d}^{2}B(z)}{{\rm d}z^{2}}
\label{eq:convsmallb}
\end{equation}

\noindent The prefactor of $\Gamma(z)$ in Eq.\ (\ref{eq:beff}) is
-$\frac{1}{6}z^{2} + O(b)$ and therefore the effective Gouy-Chapman
length is:

\begin{equation}
b_{\rm eff}-b \cong \frac{-1}{12\pi l_{\rm B} b}
         \int_{0}^{\infty} {\rm d}z \,
         z^{2}B''(z) = \frac{-B_{\rm t}}{12\pi l_{\rm B}}\frac{1}{b}
\end{equation}

\noindent This result is independent on the specific form of
$B(z)$.

To obtain $w(z)$, the deviation of the density profile relative to
PB theory, Eq.~(\ref{eq:convsmallb}) can be substituted in
Eq.~(\ref{eq:wofz2}). Up to leading order in $b$ the following
expression is obtained:

\begin{equation}
w(z) =  \frac{1}{6\pi l_{\rm B}b}\frac{1}{z}
        \int_{0}^{z}{\rm d}z' \,
        B''(z')z'^{2} +
        \frac{1}{6\pi l_{\rm B}b}z^{2}\int_{z}^{\infty}
        {\rm d}z'\, \frac{1}{z'}
        B''(z')
\end{equation}

\noindent Using $B^{\rm app}(z)$, the approximated form of $B(z)$
presented in the previous subsection, this gives:

\begin{eqnarray}
\label{eq:densitysmallb} w(z)= \left\{ \begin{array}{lr}
{\displaystyle
      \frac{1}{6\pi l_{\rm B}b}z^{2}\left(\frac{4\pi}{3}
           +\frac{B_{0}+\pi d_{\rm hc}^{2}}{d_{\rm hc}(d_{\rm hc}+\Delta)}
           -2\pi \log \frac{d_{\rm hc}}{z} \right) } \; ,
             & \left| z \right| \le d_{\rm hc} \\ \\
      {\displaystyle
      \frac{1}{6\pi l_{\rm B}b}\left[ d_{\rm hc}^{2} \left(
         \frac{4\pi}{3}d_{\rm hc}+\frac{B_{0}+
         \pi d_{\rm hc}^{2}}{\Delta} \right) \frac{1}{z}
         \right. } \\
       \hspace{1.5in}{\displaystyle - \left.
         \frac{B_{0}+\pi d_{\rm hc}^{2}}{\Delta (d_{\rm hc}+\Delta)}
         z^{2} \right] } \; ,
          & d_{\rm hc} < \left| z \right| \le d_{\rm hc} + \Delta
      \\ \\ {\displaystyle
      \frac{B_{\rm t}}{6\pi l_{\rm B}b}\frac{1}{z} } \; ,
           & d_{\rm hc}+\Delta < \left| z \right|
     \end{array}
     \right.
\end{eqnarray}

\noindent The minimal, negative value of $w(z)$ is assumed at $z =
d_{\rm hc} + \Delta$ and is equal to:

\begin{equation}
w(d_{\rm hc}+\Delta) = \frac{B_{\rm t}}{6\pi l_{\rm B}b(d_{\rm
hc}+\Delta)} \simeq \frac{B_{\rm t}}{6\pi l_{\rm B}b d_{\rm hyd}}
\end{equation}

\noindent This results in the condition
(\ref{eq:surface_layer_limit}) for the validity of the
linearization in the surface layer limit.

Using only the quadratic term in the virial expansion is sensible
if $\int_{0}^{\infty}{\rm d}z'\,c(z')B(z-z')$ is small compared to
unity. In the surface layer limit, this integral is simply:
$(|\sigma|/e)B(z) = B(z)/(2\pi l_{\rm B}b)$. Estimating the
maximum value of $|B(z)|$ to be approximately $-B_{\rm t}/(2d_{\rm
hyd})$ we obtain the requirement: $-B_{\rm t}/(4\pi l_{\rm B} b
d_{\rm hyd}) \ll 1$ , which is analogous to
(\ref{eq:surface_layer_limit}).

\subsection{Effective Gouy-Chapman length}

Using $B^{\rm app}(z)$ in equations (\ref{eq:conv}) and
(\ref{eq:beff}) we find the following approximation for the
effective Gouy-Chapman length:

\begin{eqnarray}
\label{eq:beffapprox} b_{\rm eff} - b  & = & \frac{1}{12\pi l_{\rm
B} b} \left\{
     -B^{\rm app}_{\rm t} - \pi d_{\rm hc}^{2} b + 2\pi d_{\rm hc} b^{2}
     + 2B_{0}\log\left( \frac{b+d_{\rm hc}+\Delta}{b}\right)b
   \right. \\
\nonumber
 & &  - \frac{2}{\Delta} \left(
           \pi d_{\rm hc}^{2}\Delta + \pi d_{\rm hc}^{3} + B_{0}d_{\rm hc}
                         \right)
        \log \left(\frac{b+d_{\rm hc}}{b+d_{\rm hc}+\Delta}\right)b  \\
\nonumber
  & &  \left. -  2\pi \log \left(\frac{b+d_{\rm hc}}{b}\right) b^{3}
       \right\}
\end{eqnarray}

\noindent This expression is shown in Fig.~10 and discussed
in section~\ref{sec:analytical}.
In the limits $b \gg d_{\rm hyd}$ and $b \ll d_{\rm
hyd}$ it reduces to the asymptotic expressions
(\ref{eq:befflargeb}) and (\ref{eq:beffsmallb}), respectively.
%
%
%
%
%


\begin{multicols}{2}

\end{multicols}

\end{document}